\documentclass[11pt,a4paper]{article}
\usepackage[margin=1in]{geometry}
\usepackage[authoryear]{natbib}
\usepackage{amsmath,amssymb}
\usepackage{bbm}
\usepackage{tikz}
\usetikzlibrary{positioning, arrows.meta}
\usepackage{booktabs}
\usepackage{multirow}
\usepackage{placeins}
\usepackage{float}
\usepackage{xcolor}
\usepackage{hyperref}
\usepackage{tabularx}
\usepackage{array}
\usepackage{subcaption}
\usepackage{graphicx}
\usepackage{authblk}
\usepackage{setspace}
\usepackage{titlesec}
\usepackage{listings}
\hypersetup{colorlinks=true,linkcolor=blue!60!black,citecolor=blue!60!black,urlcolor=blue!60!black}

\newcommand{\eg}{\textit{e.g.}}

\newcommand{\bench}{\textsc{GeoAnalystBench}}
\newcommand{\sys}{GISclaw}
\title{\sys{}: A Comprehensive Open-Source LLM Agent System for Realistic Multi-Step Geospatial Analysis}

\author[1]{Jinzhen Han}
\author[1]{JinByeong Lee}
\author[2]{Yuri Shim}
\author[3,*]{Jisung Kim}
\author[4,**]{Jae-Joon Lee}

\affil[1]{Department of Civil, Architectural \& Environment Engineering, Sungkyunkwan University, Suwon, South Korea}
\affil[2]{Ministry of the Interior and Safety, South Korea}
\affil[3]{School of Geography, University of Leeds, United Kingdom}
\affil[4]{Department of Fire Safety Engineering, Jeonju University, Jeonju-si, Republic of Korea}
\affil[*]{Corresponding author: gyjki@leeds.ac.uk}
\affil[**]{Corresponding author: safety\_lee@jj.ac.kr}

\date{}

\begin{document}
\maketitle
\begin{abstract}
\noindent
Most LLM-driven GIS assistants reported to date solve narrow, single-step tasks (a buffer here, a clip there) and are tightly coupled to proprietary platforms such as ArcGIS or QGIS, limiting their usefulness for the multi-step, cross-format pipelines that define professional geospatial analysis.
We present \sys{}, a comprehensive open-source agent system that performs \emph{realistic} GIS analysis end to end---spatial joins, raster algebra, kriging interpolation, machine-learning classification, network analysis, choropleth cartography---directly through Python, with no dependency on any commercial GIS software.
\sys{} couples an LLM reasoning core with a persistent Python sandbox pre-loaded with the open-source geospatial stack (GeoPandas, rasterio, scipy, scikit-learn, libpysal), three engineered prompt rules (Schema Analysis, Package Constraint, Domain Knowledge Injection), and an Error-Memory module for self-correction.
A single backend-agnostic architecture supports both cloud-API and locally deployed open-weight LLM backends, enabling air-gapped deployment without loss of capability.
On \bench{}---50 expert-curated multi-step tasks averaging 5.8 analytical steps across vector, raster, and tabular data---\sys{} reaches up to $100\%$ task success and $97\%$ mean success over three independent runs.
We further conduct $1{,}800$ controlled experiments (50 tasks $\times$ 6 backends $\times$ 2 architectures $\times$ 3 repeated runs) with bootstrap 95\% confidence intervals, paired Wilcoxon signed-rank tests, and a composite-score sensitivity analysis (Kendall's $\tau$ median $= 0.94$), and introduce a three-layer evaluation protocol combining code structure, reasoning-process, and type-specific output verification.
The Single-Agent ReAct loop reliably outperforms the Dual-Agent Plan-Execute-Replan pipeline on every cloud backend (Cliff's $\delta = 0.15$--$0.41$); only the locally deployed 14B model gains from multi-agent orchestration, suggesting that architectural complexity should be matched to model capability rather than added by default.
\end{abstract}

\noindent\textbf{Keywords:} GIS agent system; Large language models; Realistic multi-step geospatial analysis; Open-source GIS; Code generation; Persistent Python sandbox
\FloatBarrier
\section{Introduction}\label{sec:intro}
Geographic Information Systems (GIS) underpin spatial decision-making in domains ranging from urban planning and environmental monitoring to disaster response and public health~\citep{chenzhen2025genai}.
Modern geospatial analysis demands proficiency across multiple data modalities---vector geometries for boundaries and infrastructure, raster grids for satellite imagery and elevation models, and tabular attributes for census and sensor data---using specialized toolchains such as ArcGIS, QGIS, and programmatic libraries including GeoPandas, rasterio, and scipy.
This steep technical barrier limits accessibility for domain scientists who understand \textit{what} analysis is needed but lack the programming skills to implement \textit{how}, creating a persistent bottleneck between geospatial questions and actionable answers.
The rapid advancement of Large Language Models (LLMs) with strong code generation and reasoning capabilities has opened a promising pathway to bridge this gap through \textit{autonomous GIS agents}---systems that translate natural-language instructions into executable spatial analysis workflows~\citep{li2025giscience}.
\citet{li2023autonomous} formalized this vision as ``Autonomous GIS,'' demonstrating that GPT-4 could generate and execute spatial analysis code with approximately 80\% success.
Subsequent systems advanced different axes of this vision: GIS~Copilot~\citep{akinboyewa2025giscopilot} integrates LLM-guided tool selection within QGIS (86\% success on 110 tasks); GeoGPT~\citep{geogpt2024} integrates GPT-3.5 with a GIS tool pool for autonomous spatial data collection and analysis; GeoJSON~Agents~\citep{geojsonagents2025} compare code-generation vs.\ function-calling paradigms (97\% on 70 tasks); GeoColab~\citep{geocolab2025} introduces a three-role multi-agent framework with RAG (+7--26\% over single-agent baselines); GTChain~\citep{zhang2025gtchain} fine-tunes LLaMA-2-7B on synthetic tool-use chains; and GeoCode-GPT~\citep{hou2025geocodegpt} demonstrates that domain pretraining on geospatial code corpora can outperform general-purpose code LLMs on Python-based GIS analysis.
A parallel thread of multi-agent work---including the autonomous data-retrieval framework of \citet{ning2025dataretrieval}, the multi-agent GeoQA portal of \citet{feng2025geoqa}, the RAG-augmented GeoAgentic-RAG architecture of \citet{liang2026geoagenticrag}, and the multi-agent geospatial knowledge-base question-answering system of \citet{yang2026geokbqa}---has explored role decomposition, retrieval grounding, and structured-knowledge integration as routes to richer geospatial reasoning, although none of these systems isolates the contribution of multi-agent orchestration through controlled comparison against a strong single-agent baseline.
On the evaluation side, \bench{}~\citep{geoanalystbench2025} provides the most comprehensive GIS benchmark to date (50 expert tasks), complemented by GeoBenchX~\citep{geobenchx2025} for multi-step geospatial reasoning, the multi-task evaluation of \citet{xu2025evalgeo}, the JavaScript-focused GeoJSEval framework of \citet{chen2025geojseval}, and cloud-based benchmarks~\citep{cbgb2025}, while metrics research has produced CodeBLEU~\citep{ren2020codebleu}, LLM-as-a-Judge~\citep{zheng2024judging}, and embedding-based similarity~\citep{reimers2019sentence}---each capturing complementary aspects of agent quality.
Despite these advances, three critical limitations persist across existing systems, all rooted in \textbf{system design} rather than model capability:
\textbf{(1)~Narrow data-type coverage.}
Although some systems incorporate basic raster support (e.g., GIS~Copilot via QGIS raster tools, GTChain via Kriging/density operators), none provides an integrated workflow that spans vector overlays, raster interpolation, spectral index computation, and machine learning (clustering, classification) within a single agent execution.
\textbf{(2)~Platform dependency and limited model diversity.}
Most systems are tightly coupled to a single LLM (typically GPT-4) and often depend on proprietary GIS platforms~\citep{li2023autonomous,akinboyewa2025giscopilot}.
This coupling prevents systematic comparison across model families and precludes deployment in air-gapped or resource-constrained environments where API access is unavailable.
\textbf{(3)~Fragmented evaluation methodology.}
Existing metrics---binary task success, CodeBLEU, or single-dimension LLM judging---each capture only one facet of agent quality~\citep{geocodeeval2025,gramacki2024evaluation}, and no prior work unifies them into a coherent multi-layer protocol.
The functional equivalence problem in GIS code---where a buffer analysis can be correctly implemented via \texttt{gpd.overlay()}, \texttt{shapely.buffer()}, or \texttt{gpd.sjoin()}, all producing identical outputs---further renders lexical metrics unreliable.
To address these system-level gaps, we present \sys{}, a comprehensive open-source LLM agent system designed to perform \textbf{realistic, multi-step geospatial analysis} of the kind a professional GIS analyst would carry out by hand---kriging-based hot-spot mapping, Random-Forest mineral prospectivity prediction, multi-criteria wildlife corridor optimization, network-based travel-time isochrones---directly through the open-source Python ecosystem, with no dependency on any commercial GIS software.
\sys{} couples a persistent Python sandbox pre-loaded with the open-source geospatial stack (GeoPandas, rasterio, scipy, scikit-learn, libpysal, geoplot, cartopy) to a model-agnostic LLM reasoning core; three engineered prompt rules (Schema Analysis, Package Constraint, Domain Knowledge Injection) and an Error-Memory module bridge the recurring task--data information gap and enable iterative self-correction.
As Table~\ref{tab:comparison} summarizes, \sys{} differs from existing GIS agent systems along several axes simultaneously, with the most important being the realism and analytical depth of the supported workflows: we cover all 50 expert-curated multi-step tasks of \bench{}~\citep{geoanalystbench2025}---averaging 5.8 analytical steps each and spanning vector, raster, and tabular data---rather than the simpler single-tool clip/buffer pipelines that dominate prior systems.
Our contributions are:
\begin{itemize}
    \item \textbf{A comprehensive open-source agent system for realistic GIS analysis.} \sys{} is, to our knowledge, the first LLM-driven GIS agent that runs end-to-end professional analytical pipelines---spatial overlays, raster algebra, geostatistical interpolation, machine-learning classification, network analysis, and publication-quality cartography---entirely on the open-source Python stack, without falling back to ArcPy, QGIS plugins, or any other proprietary GIS layer. The system pairs a persistent Jupyter-like sandbox with a backend-agnostic LLM interface; we release the system, the prompt templates, and the open-source-rewritten gold-standard solutions for all 50 \bench{} tasks under an MIT licence.
    \item \textbf{Demonstrated coverage of professional-grade GIS workflows.} On the 50 multi-step \bench{} tasks---each modelled on a real-world analysis (urban heat-island and at-risk population mapping, mineral prospectivity prediction, flood risk under sea-level rise, fire-station coverage gap analysis, wildlife corridor optimization, etc.)---\sys{} achieves up to $100\%$ task success and $97\%$ mean success over three repeated runs, with realistic outputs that include georeferenced raster surfaces, multi-layer choropleths, and cartographically complete maps rather than only intermediate computations.
    \item \textbf{A three-layer evaluation protocol for GIS agents.} Existing benchmarks score agents by single dimensions---binary task success, code-similarity metrics, or LLM-as-Judge ratings---each capturing only a facet of agent quality. We propose a unified protocol combining (i) code structure analysis (CodeBLEU + API~Operation~F1), (ii) reasoning-process assessment (embedding similarity, complemented by an LLM-as-Judge as a qualitative diagnostic), and (iii) type-specific output verification (multimodal vision scoring for visualizations, programmatic raster/vector/tabular comparison). Composite scoring under this protocol is shown to be robust under sensitivity analysis of the dimension weights (median Kendall's $\tau = 0.94$).
    \item \textbf{Methodological validation via $1{,}800$ controlled experiments.} Across 50 tasks $\times$ 6 backends $\times$ 2 architectures $\times$ 3 repeated runs, statistical analysis (95\% bootstrap CIs, paired Wilcoxon, Cliff's $\delta$) supports two findings that we believe will be useful to the broader GIS-agent community: (i) per-task performance varies non-trivially across runs even for flagship models, motivating repeated-run protocols as a default; and (ii) within \sys{}'s identical infrastructure, the Single-Agent ReAct loop reliably outperforms a Dual-Agent Plan-Execute-Replan pipeline on every cloud backend, with only the locally deployed 14B model gaining from multi-agent orchestration---suggesting that architectural complexity should be matched to model capability rather than added by default.
\end{itemize}
The remainder of this paper is organized as follows:
Section~\ref{sec:method} details the system architecture and methodology;
Section~\ref{sec:eval} introduces the evaluation framework and benchmark;
Section~\ref{sec:experiments} presents experimental results and design analysis;
Section~\ref{sec:conclusion} concludes with future directions.
\begin{table*}[!htbp]
\centering
\caption{Comparison of \sys{} with existing LLM-powered GIS agent systems.}
\label{tab:comparison}
\resizebox{\textwidth}{!}{%
\begin{tabular}{l cccccc}
\toprule
\textbf{System} & \textbf{Data Types} & \textbf{Models} & \textbf{Architectures} & \textbf{Open-Source} & \textbf{Tasks} & \textbf{Best Success} \\
\midrule
LLM-Geo~\citep{li2023autonomous}         & Vector     & 1 (GPT-4)      & Single     & \checkmark & Case study & $\sim$80\% \\
GIS Copilot~\citep{akinboyewa2025giscopilot} & Vector+Raster  & 2 (GPT-4/3.5)  & Single     & \checkmark & 110 & 86.4\% \\
GeoGPT~\citep{geogpt2024}                & Vector     & 1 (GPT-3.5)    & Single     & No      & Case study & --- \\
GeoJSON Agents~\citep{geojsonagents2025}  & GeoJSON    & Multiple     & Multi (2-role) & No      & 70 & 97.1\% \\
GeoColab~\citep{geocolab2025}             & Vector     & 7              & Multi (3-role) & \checkmark & 50 & +26.1\% \\
GTChain~\citep{zhang2025gtchain}          & Vector+Raster & 2 (LLaMA/GPT-4) & Single (Fine-tuned) & \checkmark & Custom & +32.5\%$^\ddagger$ \\
\midrule
\textbf{\sys{} (ours)}                   & \textbf{Vec+Ras+Tab} & \textbf{6} & \textbf{SA + DA} & \textbf{\checkmark} & \textbf{50} & \textbf{100\%} \\
\bottomrule
\end{tabular}}
\vspace{1mm}
\raggedright\footnotesize{$^\ddagger$Relative improvement over GPT-4 on the GTChain benchmark, not an absolute success rate.}
\end{table*}
\FloatBarrier
\section{Methodology}\label{sec:method}
\begin{figure*}[!htbp]
\centering
\includegraphics[width=\textwidth]{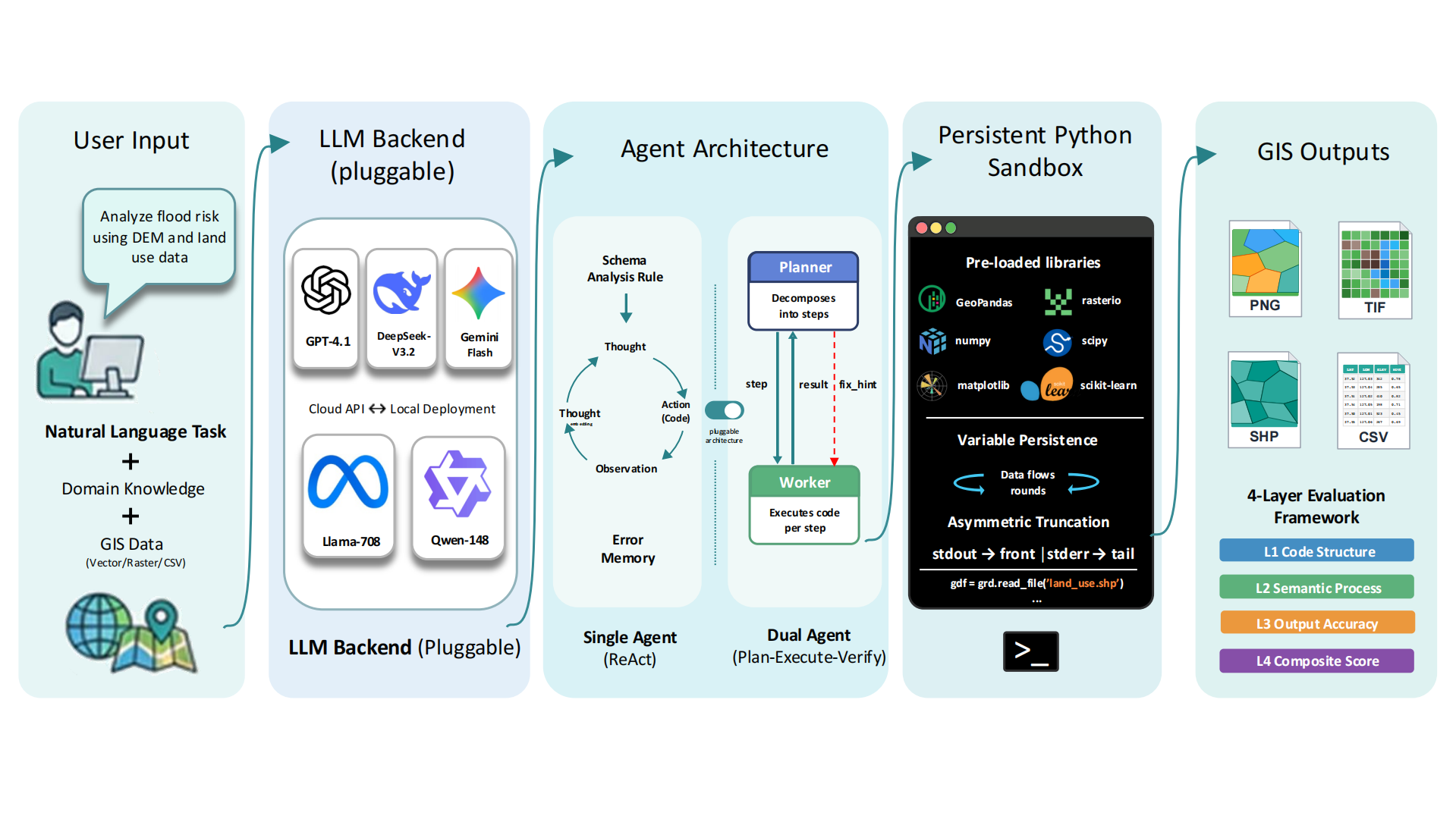}
\caption{Overview of the \sys{} system architecture. The system accepts natural-language tasks with GIS data (vector, raster, tabular) as input, routes them through a pluggable LLM backend, and executes analysis in a persistent Python sandbox. Two agent architectures---Single Agent (ReAct) and Dual Agent (Plan-Execute-Replan)---are supported, with outputs evaluated via a three-layer protocol.}
\label{fig:architecture}
\end{figure*}

\sys{} is designed around three principles:
(1)~\textit{Platform independence}: no dependency on proprietary GIS software, relying exclusively on open-source Python libraries;
(2)~\textit{Model agnosticism}: a pluggable LLM interface supporting cloud APIs and locally deployed models with identical workflows;
(3)~\textit{Domain-aware prompting}: engineered prompt rules that inject GIS-specific knowledge to compensate for LLMs' limited geospatial training data.
Fig.~\ref{fig:architecture} illustrates the overall system design.
\FloatBarrier
\subsection{System Design}\label{sec:sysdesign}
A fundamental design decision in \sys{} is its \textbf{complete independence from proprietary GIS platforms}.
Unlike systems that embed LLMs within existing GIS software---\eg, GIS~Copilot~\citep{akinboyewa2025giscopilot} couples with QGIS processing tools, and GeoGPT~\citep{geogpt2024} relies on an external GIS tool pool---\sys{} constructs its analytical capabilities entirely from open-source Python libraries.
This design choice is motivated by the observation that platform-dependent agents inherit the limitations and licensing constraints of their host software, restricting reproducibility and deployment flexibility.
Moreover, pilot experiments with non-agent, single-pass LLM code generation revealed critical deficiencies---models frequently produced syntactically valid but semantically incorrect GIS code (\eg, wrong CRS transforms, inverted raster band indices) that could not self-correct without iterative execution feedback, confirming the necessity of an agent-based architecture with sandbox integration.
By building on a self-contained Python ecosystem (Table~\ref{tab:libstack}), \sys{} ensures that any analysis expressible in Python---from basic vector overlays to machine learning-based spatial prediction---can be executed without external software dependencies.
\begin{table}[H]
\centering
\caption{Pre-loaded open-source library stack organized by analytical capability. This ecosystem replaces the need for proprietary GIS platforms.}
\label{tab:libstack}
\small
\begin{tabular}{p{1.8cm} p{2.4cm} p{3.2cm}}
\toprule
\textbf{Category} & \textbf{Libraries} & \textbf{Capabilities} \\
\midrule
Vector & GeoPandas, shapely, fiona, pyproj & Spatial joins, buffer, overlay, CRS transforms \\
\midrule
Raster & rasterio, xarray, rasterstats & Band algebra, interpolation, zonal statistics \\
\midrule
Analysis & scipy, scikit-learn, libpysal & Kriging, clustering, regression, spatial autocorrelation \\
\midrule
Visualization & matplotlib, seaborn, geoplot, cartopy & Choropleth, heatmaps, multi-panel cartography \\
\bottomrule
\end{tabular}
\end{table}
The execution core of \sys{} is a \textbf{persistent Python sandbox} that provides an interactive, Jupyter-like environment where variables and imported libraries persist across execution rounds within a single task.
Unlike stateless code-generation approaches that execute monolithic scripts, the sandbox enables the agent to execute incremental code snippets, inspect intermediate results, and iteratively refine its approach---a capability critical for exploratory GIS workflows where each analytical step depends on the results of the previous one.
The sandbox pre-loads a comprehensive GIS library stack (GeoPandas, rasterio, numpy, scipy, matplotlib, shapely, scikit-learn) upon initialization, eliminating common \texttt{ImportError} failures.
A key design choice is an \textbf{asymmetric output truncation policy}: standard output is truncated from the \textit{front} (preserving the most recent data inspection results), while error output is truncated from the \textit{tail} (preserving the root-cause error message).
This strategy is informed by the observation that GIS agents primarily need the latest intermediate results for decision-making, whereas debugging requires the original error traceback.
The sandbox further tracks new variables created in each execution round and enforces a 10-minute per-task timeout to prevent runaway computations.
Table~\ref{tab:sandbox} summarizes the key design features of the sandbox and their rationale.
\begin{table}[H]
\centering
\caption{Key design features of the persistent Python sandbox.}
\label{tab:sandbox}
\small
\begin{tabular}{p{2.2cm} p{5.2cm}}
\toprule
\textbf{Feature} & \textbf{Design Rationale} \\
\midrule
State persistence & Variables and imports survive across rounds, enabling incremental exploratory analysis \\
\midrule
Asymmetric truncation & stdout: front-truncated (latest results); stderr: tail-truncated (root-cause error) \\
\midrule
Variable tracking & New variables logged per round, providing the agent with execution context \\
\midrule
10-min timeout & Prevents runaway computations (\eg, large-scale Kriging on consumer GPUs) \\
\bottomrule
\end{tabular}
\end{table}
A central design goal of \sys{} is \textbf{model agnosticism}.
The system abstracts all LLM interaction through a unified \texttt{LLMEngine} interface, decoupling the agent logic from any specific model provider.
This interface supports cloud-hosted APIs, locally served open-weight models, and even quantized models running on a single consumer GPU, enabling deployment scenarios ranging from full cloud to fully air-gapped environments.
In practice, we found that different model families exhibit substantial variation in output formatting conventions---\eg, some models wrap code in markdown fences while others emit raw code, and reasoning-specialized models allocate internal ``thinking tokens'' that reduce the effective output budget.
\sys{} addresses these incompatibilities through provider-specific adapters within the unified interface, allowing any conforming LLM backend to be plugged in without modifying the agent logic.
\FloatBarrier
\subsection{Domain-Specific Prompt Engineering}\label{sec:prompts}
Through iterative development and failure analysis, we identified three prompt engineering rules that are critical for reliable GIS agent performance.
The first rule, \textbf{Schema Analysis}, addresses a fundamental information gap between task descriptions and actual data schemas.
GIS datasets frequently use abbreviated or domain-specific column names (\eg, ``ACSHHBPOV'' for poverty rate) that LLMs cannot infer from task descriptions alone.
To bridge this gap, the system prompt mandates that the agent's \textbf{first action} must be a data inspection step---printing column names, data types, coordinate reference systems, and sample rows---before generating any analytical code.
In ablation experiments, this single rule eliminated column-name guessing errors and significantly improved output quality for the locally deployed 14B model.
The second rule, \textbf{Package Constraint}, prevents the agent from generating code that depends on proprietary or unavailable libraries.
LLMs trained on GIS corpora frequently produce \texttt{arcpy} code (requiring an ArcGIS license) or reference packages not installed in the sandbox (\eg, \texttt{pykrige}, \texttt{skimage}).
The prompt explicitly redirects the agent to open-source equivalents such as \texttt{geopandas}, \texttt{rasterio}, and \texttt{scipy.interpolate}.
The third rule, \textbf{Optional Domain Knowledge Injection}, exposes a small input field through which a domain expert can supply task-specific procedural knowledge that lies beyond the LLM's parametric capabilities---a calibrated parameter, a non-obvious step ordering, a domain-specific threshold. When supplied, this hint is injected directly into the agent prompt; when absent, the agent proceeds with its own decomposition. The field is optional precisely because real GIS analysts sometimes have such knowledge to share and sometimes do not, and we show in Section~\ref{sec:ablation} that the system delivers comparable end-to-end task success in either mode.
Fig.~\ref{fig:prompt_example} illustrates a simplified excerpt of the system prompt, showing how these three rules are operationalized.
\begin{figure}[!htbp]
\begin{lstlisting}[caption={Simplified excerpt of the GISclaw system prompt (Single Agent mode).}, label=fig:prompt_example, captionpos=b]
You are a GIS analyst agent. Solve geospatial
analysis tasks step-by-step using tools.
## Basic Guidelines
1. Start by calling list_files to see available data.
2. Load and inspect data (print columns, CRS,
   shape, head) BEFORE writing analysis code.
3. If unsure about a library API or GIS method,
   call search_docs to look it up.
4. Save outputs to pred_results/ before finish.
5. NEVER use plt.show(). Always use
   plt.savefig(..., dpi=150, bbox_inches='tight')
## Available Packages (ONLY use these)
geopandas, rasterio, shapely, numpy, pandas,
scipy, matplotlib, sklearn, xarray, ...
NOT available (do NOT import): pykrige, arcpy
Alternatives: pykrige -> scipy.interpolate
\end{lstlisting}
\end{figure}
\FloatBarrier
\subsection{Agent Architectures}\label{sec:arch}
\begin{figure}[!htbp]
\centering
\includegraphics[width=\columnwidth]{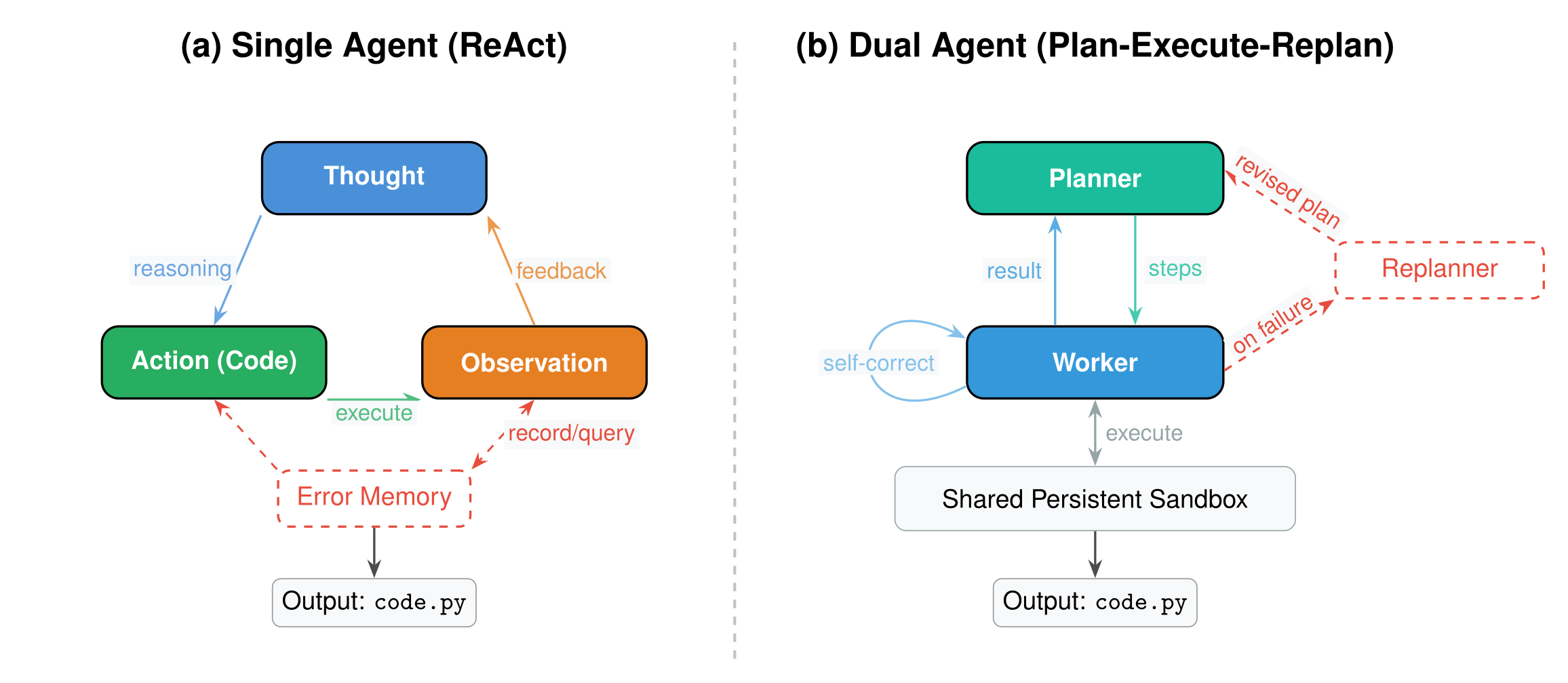}
\caption{Comparison of the two agent architectures. (a)~Single Agent follows a ReAct loop with Error Memory for self-correction. (b)~Dual Agent decomposes tasks via a Planner, executes steps through a Worker, and adaptively replans upon failure.}
\label{fig:agent_arch}
\end{figure}

\sys{} implements two pluggable agent architectures that operate on the same sandbox and prompt infrastructure, enabling controlled comparison of their effectiveness across different model capabilities.
Fig.~\ref{fig:agent_arch} illustrates the two architectures.
\textbf{Single Agent (SA)} follows the ReAct~\citep{yao2023react} paradigm, iteratively generating Thought (natural-language reasoning), Action (Python code for the sandbox), and receiving Observation (execution output or error messages).
An \texttt{ErrorMemory} module records error patterns encountered across rounds, preventing the agent from repeating failed approaches and promoting progressive refinement.
The agent accumulates all generated code into a consolidated script upon task completion.
\textbf{Dual Agent (DA)} adopts a Plan-Execute-Replan pipeline with explicit task decomposition.
A \textit{Planner} receives the task instruction, domain knowledge, and data schema (obtained via Schema Analysis), then decomposes the task into 3--7 ordered analytical steps.
A \textit{Worker} executes each step within the shared persistent sandbox, with up to 10 self-correction rounds per step.
When the Worker fails on a step, the Planner is re-invoked as a \textit{Replanner} with the failure context to generate a simplified alternative plan that avoids the failed operations; up to 2 replanning cycles are permitted.
Both roles share the same underlying LLM and sandbox namespace, ensuring information continuity.
Notably, the Replanner is not a separate module but the same Planner invoked with failure-aware context---this keeps the architecture lightweight while enabling adaptive recovery.
\FloatBarrier
\section{Evaluation Framework and Benchmark}\label{sec:eval}
\FloatBarrier
\subsection{Benchmark Dataset}
We evaluate \sys{} on \bench{}~\citep{geoanalystbench2025}, the most comprehensive publicly available benchmark for GIS agent systems.
\bench{} comprises 50 expert-designed tasks organized into six analytical categories (Table~\ref{tab:benchmark}), spanning three data modalities (vector, raster, tabular) with an average of 5.8 workflow steps per task.
Each task is modeled after a real-world GIS analysis scenario---including flood risk assessment, mineral prospectivity mapping, urban heat island analysis, and wildlife corridor optimization---requiring multi-step spatial reasoning, cross-format data integration, and domain-specific parameter calibration.
This design prioritizes analytical depth over task count: the 50 tasks collectively cover the full spectrum of geospatial operations (overlay, buffer, interpolation, raster algebra, machine learning, network analysis) at a complexity level representative of professional GIS workflows, making the benchmark a rigorous testbed for agent capability despite its moderate size.
Across all evaluation experiments, we uniformly inject the ``Human Designed Workflow'' provided by the benchmark as domain knowledge (see Section~\ref{sec:prompts}) directly into the evaluation prompt. This setup not only ensures fair comparability among models under identical problem-solving logic but also effectively isolates the vast variance introduced by unconstrained free exploration, thereby truly reflecting the Agent's ability to comprehend and execute precise professional instructions.
\textbf{Gold standard adaptation.}
A critical contribution of our work is the systematic adaptation of \bench{}'s gold-standard solutions.
The original benchmark provides expert-authored reference code written in ArcPy---a Python API tightly coupled with the proprietary ArcGIS platform.
Since \sys{} operates entirely on open-source libraries, we manually rewrote all 50 gold-standard solutions using the open-source GIS ecosystem (GeoPandas, rasterio, scipy, scikit-learn), ensuring functional equivalence while eliminating proprietary dependencies.
These rewritten solutions serve as the reference for all evaluation layers (L1--L3).
\begin{table}[H]
\centering
\caption{\bench{} task categories with representative examples. Each task includes natural-language instructions, heterogeneous input data, optional domain knowledge, and expert-authored gold-standard code.}
\label{tab:benchmark}
\small
\begin{tabular}{p{2.4cm} r p{4.4cm}}
\toprule
\textbf{Category} & \textbf{$N$} & \textbf{Representative Task} \\
\midrule
Understanding spatial distributions & 19 & T1: Urban heat island \& elderly risk mapping via Kriging interpolation \\
\midrule
Making predictions & 8 & T20: Mineral deposit prediction using Random Forest with raster features \\
\midrule
Detecting patterns & 7 & T42: Airbnb price clustering via local Moran's~I spatial autocorrelation \\
\midrule
Measuring shape \& distribution & 7 & T36: Vegetation change detection using SAVI spectral index \\
\midrule
Determining spatial relationships & 6 & T38: Travel-time isochrone computation on road networks \\
\midrule
Optimal locations \& paths & 3 & T32: Wildlife corridor optimization via weighted cost-surface overlay \\
\bottomrule
\end{tabular}
\end{table}
\FloatBarrier
\subsection{LLM Backends}\label{sec:llm_backends}
To validate model agnosticism, we evaluate \sys{} with six heterogeneous LLM backends spanning cloud APIs and local deployments (Table~\ref{tab:models}).
Cloud models include two OpenAI models (GPT-5.4 and GPT-4.1), DeepSeek-V3.2 (a cost-effective alternative at significantly lower price), and Google's Gemini-3-Flash.
For offline deployment, we test Llama-3.3-70B on a multi-GPU server and Qwen2.5-Coder-14B on a single consumer RTX~3090 GPU.
This selection covers the spectrum from flagship reasoning models to lightweight code-specialized models, enabling systematic comparison of how model capability interacts with system design.
\begin{table}[H]
\centering
\caption{LLM backends evaluated. Cost is per million tokens (input/output). Open-weight models are deployed on local servers at zero marginal cost.}
\label{tab:models}
\small
\begin{tabular}{llrr}
\toprule
\textbf{Model} & \textbf{Backend} & \textbf{Cost (\$/M)} & \textbf{Context} \\
\midrule
\multicolumn{4}{l}{\textit{Cloud API (pay-per-use)}} \\
GPT-5.4        & OpenAI    & 2.50 / 15.00  & 1M \\
GPT-4.1        & OpenAI    & 2.00 / 8.00  & 1M \\
DeepSeek-V3.2  & DeepSeek  & 0.28 / 0.42  & 128K \\
Gemini-3-Flash & Google AI & 0.50 / 3.00  & 1M \\
\midrule
\multicolumn{4}{l}{\textit{Local deployment (zero marginal cost)}} \\
Llama-3.3-70B  & Local server  & Free  & 128K \\
Qwen2.5-14B    & Local GPU     & Free  & 32K \\
\bottomrule
\end{tabular}
\end{table}
\FloatBarrier
\subsection{Multi-Layer Evaluation Protocol}
To comprehensively assess agent performance beyond binary success rates, we propose a multi-layer evaluation protocol that captures complementary aspects of agent quality---from surface-level code fidelity to deep reasoning process and final output correctness (Fig.~\ref{fig:eval_pipeline}).
Individual layer scores are combined into a composite score:
\begin{equation}
    S_{\text{comp}} = 0.4 \cdot R_{\text{succ}} + 0.3 \cdot Q_{\text{out}} + 0.15 \cdot F_{\text{api}} + 0.15 \cdot C_{\text{emb}}
    \label{eq:composite}
\end{equation}
where $Q_{\text{out}}$ averages only over \textit{successful} tasks to avoid double-penalizing failures already captured by $R_{\text{succ}}$.
\begin{figure}[!htbp]
\centering
\resizebox{\columnwidth}{!}{%
\begin{tikzpicture}[
    node distance=0.3cm,
    layer/.style={draw, rounded corners, minimum width=2.8cm, minimum height=0.8cm, font=\small\bfseries, text=white},
    metric/.style={draw, rounded corners=2pt, minimum width=2.2cm, minimum height=0.55cm, font=\scriptsize, fill=white},
    arrow/.style={->, thick, >=stealth}
]
\node[layer, fill=blue!70] (l1) {L1: Code};
\node[metric, right=0.6cm of l1, yshift=0.45cm] (cb) {CodeBLEU};
\node[metric, right=0.6cm of l1, yshift=-0.45cm] (api) {API Op.\ F1};
\draw[arrow] (l1.east) -- (cb.west);
\draw[arrow] (l1.east) -- (api.west);
\node[layer, fill=green!60!black, below=1.2cm of l1] (l2) {L2: Process};
\node[metric, right=0.6cm of l2, yshift=0.45cm] (emb) {Embedding Sim.};
\node[metric, right=0.6cm of l2, yshift=-0.45cm] (judge) {LLM-as-Judge};
\draw[arrow] (l2.east) -- (emb.west);
\draw[arrow] (l2.east) -- (judge.west);
\node[layer, fill=orange!80, below=1.2cm of l2] (l3) {L3: Output};
\node[metric, right=0.6cm of l3, yshift=0.9cm] (png) {PNG: Vision};
\node[metric, right=0.6cm of l3, yshift=0.0cm] (tif) {TIF: Rasterio};
\node[metric, right=0.6cm of l3, yshift=-0.9cm] (csv) {CSV/SHP};
\draw[arrow] (l3.east) -- (png.west);
\draw[arrow] (l3.east) -- (tif.west);
\draw[arrow] (l3.east) -- (csv.west);
\node[layer, fill=red!70, minimum width=3.6cm, minimum height=5.5cm, right=4.5cm of l2, yshift=-0.2cm] (comp) {\Large $S_{\text{comp}}$};
\draw[arrow] (cb.east) -- (comp.west |- cb.east);
\draw[arrow] (api.east) -- (comp.west |- api.east);
\draw[arrow] (emb.east) -- (comp.west |- emb.east);
\draw[->, thick, >=stealth, dashed, gray] (judge.east) -- node[above, font=\tiny, text=gray, fill=white, inner sep=1pt] {qualitative} (comp.west |- judge.east);
\draw[arrow] (png.east) -- (comp.west |- png.east);
\draw[arrow] (tif.east) -- (comp.west |- tif.east);
\draw[arrow] (csv.east) -- (comp.west |- csv.east);
\end{tikzpicture}%
}
\caption{Multi-layer evaluation pipeline. Three complementary layers assess code-level fidelity (L1), reasoning process quality (L2), and output correctness (L3), combined into a weighted composite score $S_{\text{comp}}$ (Eq.~\ref{eq:composite}).}
\label{fig:eval_pipeline}
\end{figure}
\textbf{L1: Code Structure.}
We evaluate syntactic and structural similarity between generated and gold code using four metrics.
\textit{BLEU-4}~\citep{papineni2002bleu} computes geometric-mean $n$-gram precision ($n=1{\ldots}4$) with brevity penalty:
\begin{equation}
    \text{BLEU-4} = \text{BP} \cdot \exp\!\Bigl(\tfrac{1}{4}\textstyle\sum_{n=1}^{4}\log p_n\Bigr)
    \label{eq:bleu}
\end{equation}
where $p_n$ denotes clipped $n$-gram precision and $\text{BP}=\min(1, e^{1-|r|/|c|})$.
\textit{ROUGE-L} measures the longest common subsequence (LCS):
\begin{equation}
    R_{\text{lcs}} = \tfrac{|{\rm LCS}|}{|r|},\quad
    P_{\text{lcs}} = \tfrac{|{\rm LCS}|}{|c|},\quad
    F_{\text{lcs}} = \tfrac{2 R_{\text{lcs}} P_{\text{lcs}}}{R_{\text{lcs}}+P_{\text{lcs}}}
    \label{eq:rouge}
\end{equation}
\textit{CodeBLEU}~\citep{ren2020codebleu} extends BLEU with code-aware components:
\begin{equation}
    \text{CodeBLEU} = \tfrac{1}{4}\bigl(\alpha_{\text{ngram}} + \alpha_{\text{wt}} + \alpha_{\text{syn}} + \alpha_{\text{df}}\bigr)
    \label{eq:codebleu}
\end{equation}
where $\alpha_{\text{ngram}}$ is standard BLEU-4, $\alpha_{\text{wt}}$ is keyword-weighted $n$-gram match (Python keywords receive double weight), $\alpha_{\text{syn}}$ is AST subtree match F1, and $\alpha_{\text{df}}$ is data-flow match F1 based on variable define--use chains.
We additionally compute \textit{Edit Similarity} as $1 - d_{\text{Lev}} / \max(|r|,|c|)$.
Critically, we introduce \textbf{API Operation~F1}: extracting GIS-specific operations (\eg, \texttt{spatial\_join}, \texttt{buffer}, \texttt{overlay}, \texttt{kriging}) from both generated and reference code via pattern matching, and computing set-level precision, recall, and F1.
This metric is robust to the functional equivalence problem, where semantically correct implementations share minimal lexical overlap.
\textbf{L2: Reasoning Process.}
This layer evaluates whether the agent's analytical \textit{reasoning process} aligns with expert expectations, using two complementary methods.
First, we encode cleaned execution logs and gold references using OpenAI's \texttt{text-embedding-3-large} model (3072-d) and compute cosine similarity:
\begin{equation}
    C_{\text{emb}} = \frac{\mathbf{e}_{\text{agent}} \cdot \mathbf{e}_{\text{gold}}}
    {\|\mathbf{e}_{\text{agent}}\| \cdot \|\mathbf{e}_{\text{gold}}\|}
    \label{eq:cosine}
\end{equation}
This provides a continuous, reference-aligned measure of process quality.
Second, we employ an \textbf{LLM-as-Judge} protocol~\citep{zheng2024judging}: DeepSeek-V3.2 in non-thinking mode---chosen for cost-controlled evaluation across the full $1{,}800$-experiment grid and held fixed across all runs to ensure cross-run comparability---scores each execution log on five dimensions (1--5 scale):
(i)~\textit{Task Understanding}: correct interpretation and planning;
(ii)~\textit{Data Handling}: loading, CRS management, format handling;
(iii)~\textit{Methodology}: appropriateness of GIS analysis methods;
(iv)~\textit{Self-Correction}: error detection and recovery efficiency;
(v)~\textit{Result Completeness}: completeness of final deliverables.
The judge receives the task instruction, expert-authored gold code, and the complete execution log, producing per-dimension scores with natural-language justifications.
This dual assessment captures both statistical process alignment (embedding) and nuanced qualitative reasoning evaluation (LLM-as-Judge).
Of the two, only $C_{\text{emb}}$ enters the composite score (Eq.~\ref{eq:composite}); the LLM-as-Judge scores serve as a qualitative diagnostic tool for the case-study analysis in Section~\ref{sec:experiments}.
\textbf{L3: Output Accuracy.}
Generated outputs are evaluated against gold standards using \textbf{type-specific methods} tailored to the heterogeneous output formats of GIS tasks:
\textit{Visualization outputs} (PNG): a multimodal vision model (Gemini-2.5-Flash-Lite, chosen for cost control across the $1{,}800$-experiment grid and yielding $100\%$ JSON-valid responses with zero parse errors) receives both the gold and agent images alongside the task instruction, scoring five cartographic dimensions: Task Completion, Spatial Accuracy, Visual Readability, Cartographic Quality, and Data Integrity (each 1--5), explicitly accepting alternative-but-correct visualization styles. Blank or near-blank images are pre-filtered by a PIL-based heuristic ($\leq 5$ unique colours) before the vision call.
\textit{Raster outputs} (GeoTIFF): evaluated programmatically via CRS match, shape match, pixel-level Pearson correlation, and mean relative error against the gold raster, yielding a weighted composite:
$s_{\text{raster}} = 0.2 \cdot \mathbbm{1}_{\text{shape}} + 0.2 \cdot \mathbbm{1}_{\text{CRS}} + 0.3 \cdot f(\rho) + 0.3 \cdot g(\text{MRE})$, where $f$ and $g$ are piecewise thresholding functions.
\textit{Tabular outputs} (CSV): assessed through column overlap ratio, row-count match, and average Pearson correlation across shared numeric columns.
\textit{Vector outputs} (Shapefile, GeoJSON, GeoPackage): compared via feature count match, CRS consistency, and column overlap ratio against the gold reference.
All type-specific scores are normalized to $[0, 1]$ (vision scores are divided by the maximum scale of~5) and averaged across output files to yield a per-task output accuracy score $Q_{\text{out}}$.
\FloatBarrier
\section{Experimental Results and Analysis}\label{sec:experiments}
To evaluate \sys{} thoroughly, we run the system on every \bench{} task across the six LLM backends introduced in Section~\ref{sec:llm_backends} and the two agent architectures, repeated three times per cell---$1{,}800$ controlled executions in total.
Each task is allocated a 20-minute hard timeout and a maximum of 50 interaction rounds; sampling temperature, prompts, and sandbox configuration are held fixed across runs, so any per-task variance reflects the stochasticity of the LLM samples rather than configuration drift.
The sandbox environment runs Ubuntu with Python~3.12 and the pre-loaded GIS library stack described in Section~\ref{sec:sysdesign}.
Table~\ref{tab:master} presents the aggregated results, with each cell averaged over its three repeated runs.
\sys{} reaches a peak mean success rate of $100\%$ (GPT-5.4, SA) and three further backends exceed $96\%$ mean success in SA mode, indicating that the system reliably executes the entire 50-task realistic-pipeline benchmark even with mid-tier cloud APIs and the locally deployed Llama-3.3-70B.
The Single-Agent ReAct loop is the stronger architecture for every cloud backend; the Dual-Agent Plan-Execute-Replan pipeline only outperforms its Single-Agent counterpart on the smallest model in the suite (Qwen2.5-Coder-14B), a pattern we examine in detail in Sections~\ref{sec:wilcoxon} and~\ref{sec:variance}.
The following subsections analyse each evaluation layer in turn.
\begin{table*}[t]
\centering
\caption{Comprehensive evaluation across all six LLM backends and two architectures, averaged over three repeated runs (50 tasks per cell, $n=150$ task-runs per model--architecture). $R_s$: task success rate; $Q_\text{out}$: output accuracy (L3, successful tasks only); $F_\text{api}$: API operation F1 (L1); $C_\text{emb}$: embedding cosine similarity (L2); $S_\text{comp}$: composite score (Eq.~\ref{eq:composite}) with bootstrap 95\% confidence interval. Cost is the total API expenditure averaged per benchmark sweep (50 tasks, single run). Bold = best per architecture.}
\label{tab:master}
\small
\resizebox{\textwidth}{!}{%
\begin{tabular}{l cccccc cccccc}
\toprule
& \multicolumn{6}{c}{\textbf{Single Agent}} & \multicolumn{6}{c}{\textbf{Dual Agent}} \\
\cmidrule(lr){2-7} \cmidrule(lr){8-13}
\textbf{Model} & $R_s$ & $Q_\text{out}$ & $F_\text{api}$ & $C_\text{emb}$ & $\boldsymbol{S_\textbf{comp}}\ (\pm\textbf{CI})$ & \textbf{Cost} & $R_s$ & $Q_\text{out}$ & $F_\text{api}$ & $C_\text{emb}$ & $\boldsymbol{S_\textbf{comp}}\ (\pm\textbf{CI})$ & \textbf{Cost} \\
\midrule
DeepSeek-V3.2  & 0.973 & 0.635 & 0.632 & \textbf{0.641} & 0.761 (\,$\pm$0.023) & \$0.50 & 0.893 & \textbf{0.532} & \textbf{0.585} & 0.636 & \textbf{0.681 (\,$\pm$0.033)} & \$0.80 \\
Gemini-3-Flash & 0.967 & 0.635 & \textbf{0.662} & \textbf{0.647} & \textbf{0.765 (\,$\pm$0.024)} & \$0.80 & 0.867 & \textbf{0.540} & 0.532 & 0.562 & 0.650 (\,$\pm$0.036) & \$1.50 \\
GPT-5.4        & \textbf{1.000} & 0.562 & 0.602 & 0.633 & 0.750 (\,$\pm$0.016) & \$8.50 & \textbf{0.967} & 0.507 & 0.538 & 0.628 & \textbf{0.705 (\,$\pm$0.022)} & \$12.30 \\
GPT-4.1        & 0.967 & 0.601 & \textbf{0.663} & 0.604 & 0.747 (\,$\pm$0.026) & \$1.20 & 0.933 & 0.503 & 0.573 & \textbf{0.648} & 0.695 (\,$\pm$0.027) & \$3.40 \\
Llama-3.3-70B  & 0.913 & 0.398 & 0.610 & 0.614 & 0.656 (\,$\pm$0.029) & \$0$^\dagger$ & 0.893 & 0.344 & 0.577 & 0.620 & 0.627 (\,$\pm$0.030) & \$0$^\dagger$ \\
Qwen2.5-14B    & 0.520 & \textbf{0.549} & 0.518 & 0.590 & 0.454 (\,$\pm$0.050) & \$0$^\dagger$ & 0.787 & 0.427 & \textbf{0.586} & 0.599 & 0.588 (\,$\pm$0.038) & \$0$^\dagger$ \\
\bottomrule
\end{tabular}}
\vspace{1mm}
\raggedright\footnotesize{$^\dagger$Open-weight models deployed locally; zero marginal cost.}
\end{table*}
\FloatBarrier
\subsection{Quantitative Evaluation Results}
\FloatBarrier
\subsubsection{Task success rate and architecture comparison}\label{sec:wilcoxon}
\begin{figure}[!htbp]
\centering
\includegraphics[width=\columnwidth]{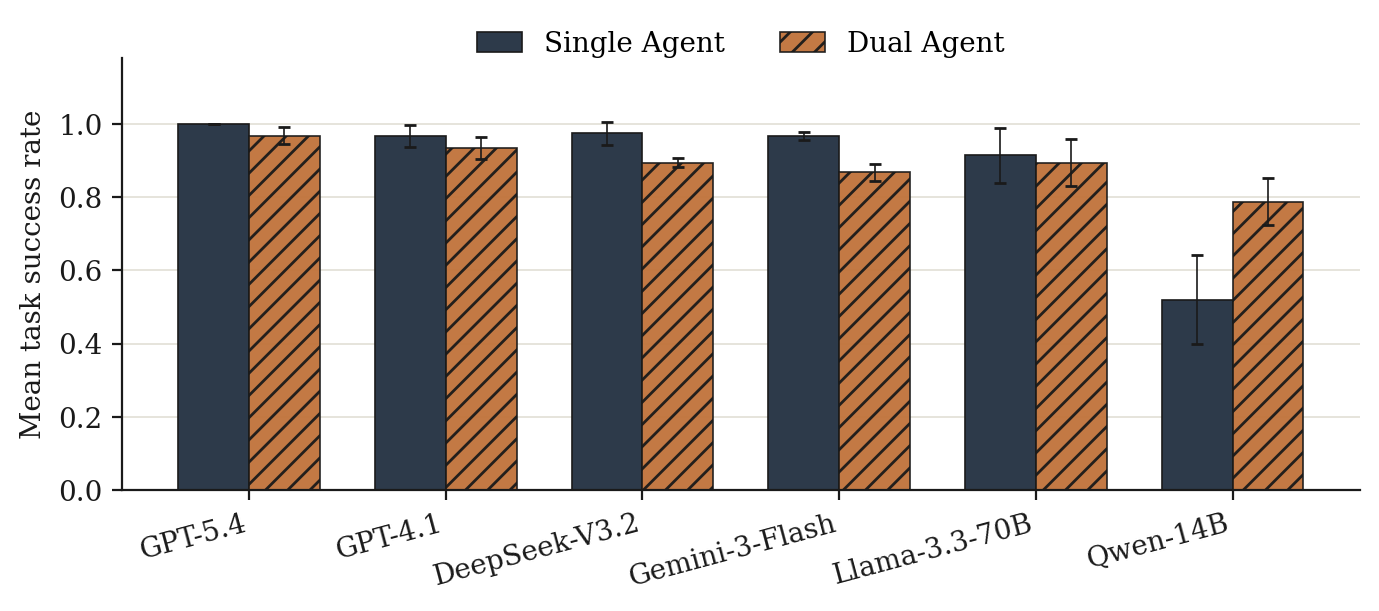}
\caption{Task success rate comparison (mean over three repeated runs): Single Agent (solid) vs.\ Dual Agent (hatched). All cloud and large open-weight models show the Single Agent reliably ahead; only Qwen2.5-Coder-14B improves under the Dual Agent.}
\label{fig:success_bar}
\end{figure}

Fig.~\ref{fig:success_bar} visualizes the SA--DA gap across all models, averaged over three repeated runs.
For all four cloud backends and the 70B open-weight model, the Single Agent reliably exceeds the Dual Agent in mean success rate by 3--10 percentage points, with the largest absolute drop on DeepSeek-V3.2 (97.3\%~$\to$~89.3\%, $-8.0$ pp) and Gemini-3-Flash (96.7\%~$\to$~86.7\%, $-10.0$ pp).
Qwen2.5-Coder-14B is the only backend whose mean success rate \emph{rises} under multi-agent orchestration (52.0\%~$\to$~78.7\%, $+26.7$ pp), consistent with the architecture--capability matching principle.
The SA--DA composite gap on the five models where SA wins is between $0.03$ and $0.11$, indicating that the Dual Agent overhead, while statistically reliable, is moderate in absolute size once measured over repeated runs at the task level rather than from any single deterministic execution.

\textbf{Statistical significance.} Table~\ref{tab:wilcoxon} reports per-model paired Wilcoxon signed-rank tests on the task-level composite scores ($n = 150$ paired task-runs per model: 50 tasks $\times$ 3 runs).
For five of six models, the Single Agent significantly outperforms the Dual Agent ($p < 0.02$ for all five; $p < 10^{-3}$ for the four cloud models), with effect sizes ranging from small (Cliff's $\delta \approx 0.15$ for Llama-3.3-70B) to medium (Gemini-3-Flash, $\delta = 0.41$).
Qwen2.5-Coder-14B is the sole reversal: the Dual Agent significantly outperforms the Single Agent ($p < 10^{-3}$, $\delta = -0.20$), confirming that the planner--worker decomposition compensates for the smaller model's weaker single-shot reasoning.

\begin{table}[H]
\centering
\caption{Paired Wilcoxon signed-rank tests on task-level composite scores: Single Agent vs.\ Dual Agent. $n=150$ paired task-runs per row (50 tasks $\times$ 3 runs); positive median difference and Cliff's $\delta$ favour SA. All but Llama-3.3-70B reach $p < 10^{-3}$.}
\label{tab:wilcoxon}
\small
\begin{tabular}{l rrr rr}
\toprule
\textbf{Model} & $\boldsymbol{S^{\text{SA}}_\text{comp}}$ & $\boldsymbol{S^{\text{DA}}_\text{comp}}$ & \textbf{Median $\boldsymbol{\Delta}$} & \textbf{Cliff's $\boldsymbol{\delta}$} & $\boldsymbol{p}$-value \\
\midrule
Gemini-3-Flash  & 0.7646 & 0.6501 & $+0.067$ & $+0.41$ & $<10^{-3}$ \\
DeepSeek-V3.2   & 0.7612 & 0.6809 & $+0.028$ & $+0.29$ & $<10^{-3}$ \\
GPT-4.1         & 0.7473 & 0.6954 & $+0.034$ & $+0.26$ & $<10^{-3}$ \\
GPT-5.4         & 0.7504 & 0.7052 & $+0.029$ & $+0.21$ & $<10^{-3}$ \\
Llama-3.3-70B   & 0.6563 & 0.6269 & $+0.010$ & $+0.15$ & $0.017$    \\
\midrule
Qwen2.5-Coder-14B & 0.4537 & 0.5882 & $-0.032$ & $-0.20$ & $<10^{-3}$ \\
\bottomrule
\end{tabular}
\end{table}

\begin{figure}[!htbp]
\centering
\includegraphics[width=\columnwidth]{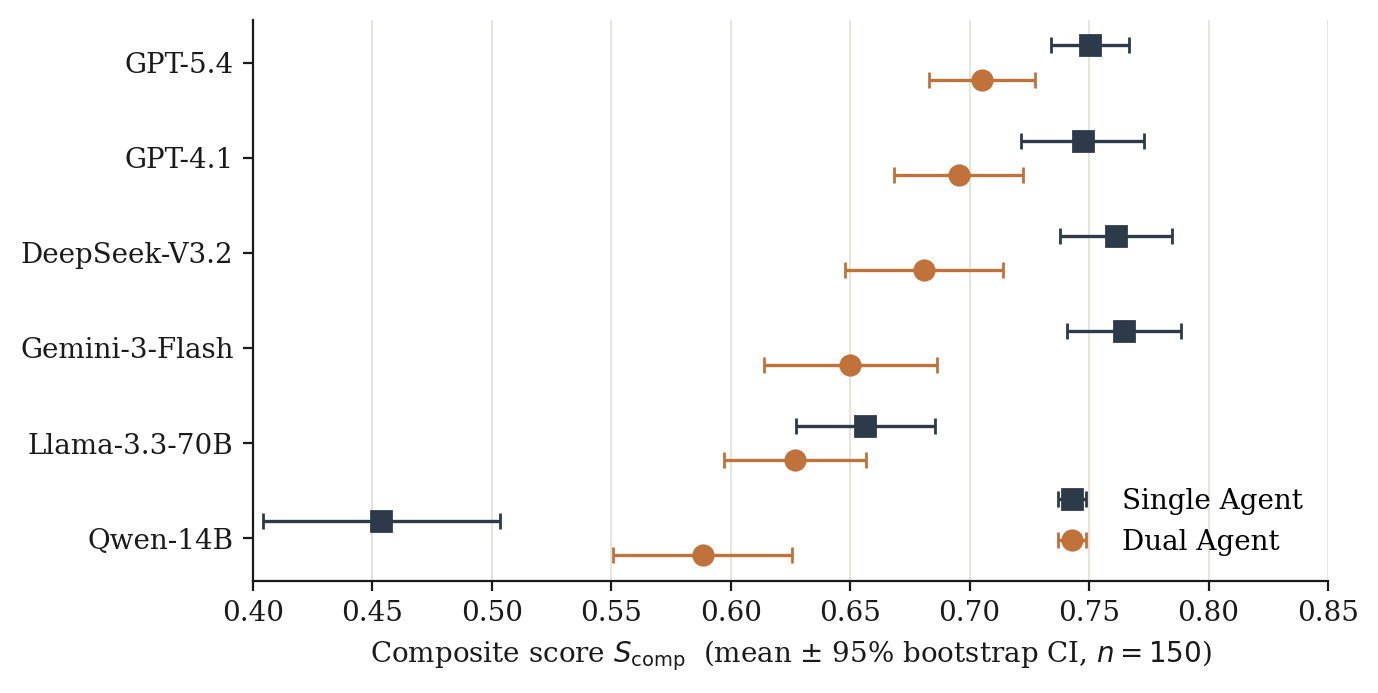}
\caption{Mean composite score $S_\text{comp}$ per (model, architecture) cell with 95\% bootstrap confidence intervals ($n=150$ paired task-runs per cell). All four cloud backends and Llama-3.3-70B sit higher under SA than DA; Qwen2.5-Coder-14B is the only backend whose CI bracket clears the corresponding SA bracket on the right.}
\label{fig:ci_errorbar}
\end{figure}

\textbf{Sensitivity to composite weights.}
Because the four-component composite (Eq.~\ref{eq:composite}) involves a choice of weights, we re-rank the twelve (model, architecture) cells under all 66 weight tuples $(w_{\text{succ}}, w_{\text{out}}, w_{\text{api}}, w_{\text{cos}})$ on the simplex $\{0.1, 0.2, 0.3, 0.4, 0.5\}$ summing to $1$.
The Kendall's $\tau$ between the resulting ranking and the paper-default ranking has median $0.94$ (range $0.79$--$1.00$), and $\tau \geq 0.90$ in 47 of 66 weight configurations.
The two qualitative conclusions---five models favour SA at the task level, and Qwen-14B favours DA---hold under every weight tuple we tested.
\FloatBarrier
\subsubsection{Code quality analysis (L1)}
\begin{figure}[!htbp]
\centering
\includegraphics[width=\columnwidth]{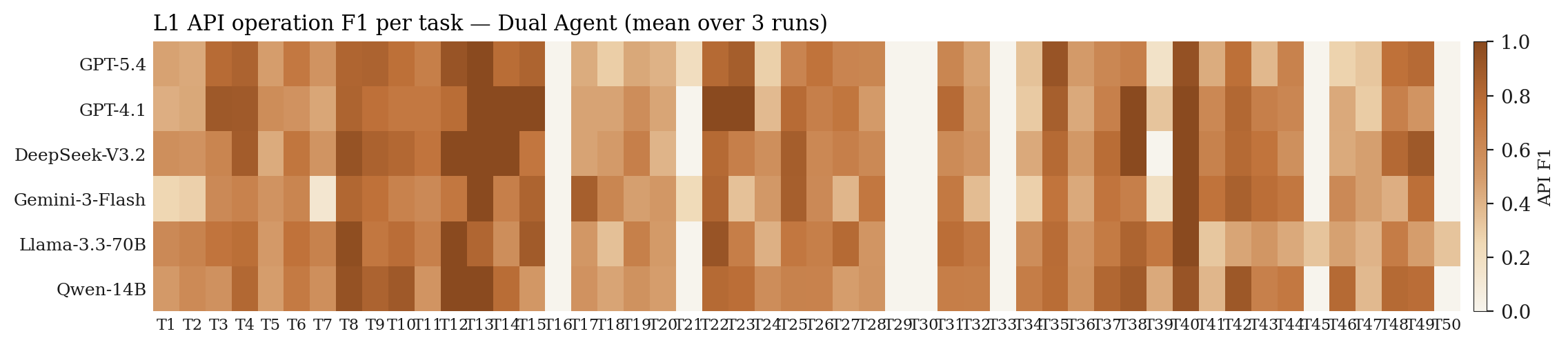}
\caption{Per-task API F1 scores (DA architecture). Warm colors indicate higher code-level agreement with gold standards. The task-dependent variance highlights the limitation of aggregate metrics.}
\label{fig:l1_heatmap}
\end{figure}

\begin{table}[H]
\centering
\caption{Code quality metrics across architectures, averaged over three repeated runs (50 tasks per cell). Bold = best per architecture.}
\label{tab:codebleu}
\small
\begin{tabular}{l ccc}
\toprule
\textbf{Model} & \textbf{CodeBLEU} & \textbf{API~F1} & \textbf{ROUGE-L} \\
\midrule
\multicolumn{4}{l}{\textit{Single Agent}} \\
GPT-5.4      & \textbf{0.172} & 0.602 & 0.136 \\
GPT-4.1      & 0.171 & 0.659 & \textbf{0.142} \\
Gemini Flash & 0.168 & \textbf{0.662} & 0.141 \\
Llama-70B    & 0.166 & 0.606 & 0.137 \\
Qwen-14B     & 0.162 & 0.504 & 0.132 \\
DeepSeek     & 0.126 & 0.632 & 0.097 \\
\midrule
\multicolumn{4}{l}{\textit{Dual Agent}} \\
Gemini Flash & \textbf{0.176} & 0.532 & \textbf{0.137} \\
GPT-4.1      & 0.163 & 0.573 & 0.132 \\
GPT-5.4      & 0.139 & 0.535 & 0.109 \\
DeepSeek     & 0.122 & 0.585 & 0.092 \\
Qwen-14B     & 0.110 & 0.574 & 0.098 \\
Llama-70B    & 0.094 & \textbf{0.577} & 0.091 \\
\bottomrule
\end{tabular}
\end{table}
A notable finding from Table~\ref{tab:codebleu}: in DA mode, Llama-70B achieves the \textit{lowest} CodeBLEU (0.094) yet near-equal API~F1 (0.577), and DeepSeek shows the same pattern (CodeBLEU $0.122$, API~F1 $0.585$). This exposes the systematic bias of CodeBLEU for GIS code---a single task (\eg, buffer analysis) can be correctly implemented via \texttt{gpd.overlay()}, \texttt{shapely.buffer()}, or \texttt{gpd.sjoin()} with manual geometry operations, all yielding equivalent results but sharing minimal lexical overlap. We recommend \textbf{API~F1} as a more reliable metric for GIS code evaluation.
Fig.~\ref{fig:l1_heatmap} presents the per-task API~F1 distribution, revealing that code similarity is highly task-dependent: simple overlay tasks (T24, T40) score consistently above 0.7 across models, while complex multi-step analyses (T27, T39) show near-zero overlap even for successful completions.
\FloatBarrier
\subsubsection{Reasoning process assessment (L2)}
\begin{figure}[!htbp]
\centering
\includegraphics[width=\columnwidth]{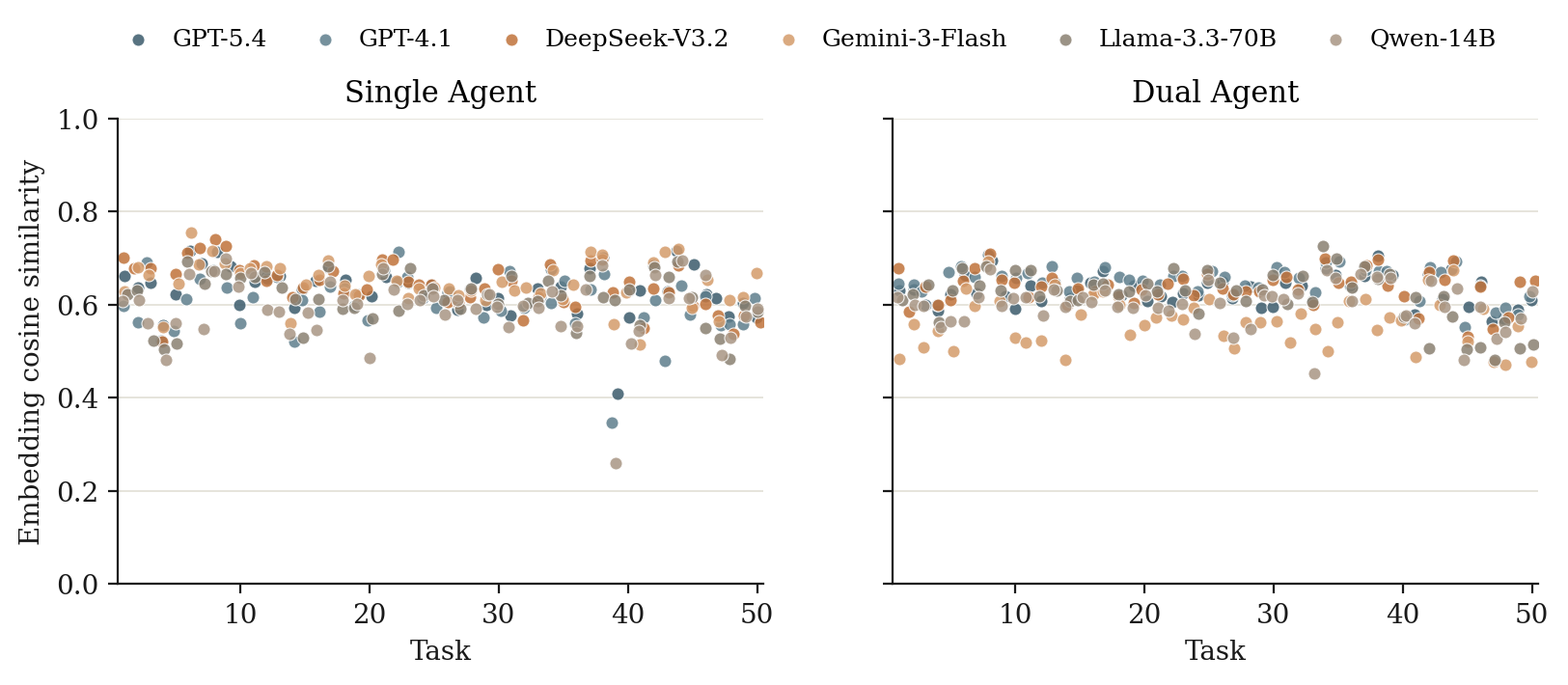}
\caption{L2: Execution log embedding cosine similarity per task (mean across three repeated runs). Each dot represents one (model, task) pair. SA and DA produce comparable mean similarity, with DA showing greater task-level variance for the same model.}
\label{fig:l2_embedding}
\end{figure}

Fig.~\ref{fig:l2_embedding} presents the execution log embedding similarity for each task--model pair, averaged over three repeated runs.
In SA mode, DeepSeek and Gemini Flash maintain consistently high cosine similarity ($\sim 0.64$ on average) across most tasks; in DA mode, the planner--worker decomposition produces an even slightly higher average similarity for some backends (\eg, GPT-4.1 SA $0.604$ vs.\ DA $0.648$) but with greater task-level dispersion.
The two architectures therefore align similarly with gold-standard reasoning on average, but the Dual Agent's planning step distributes its similarity less uniformly across tasks.
To complement the embedding-based assessment, a DeepSeek-V3.2 (non-thinking) LLM-as-Judge---deliberately distinct from the GPT- and Gemini-family models being scored---evaluates each execution log across five qualitative dimensions (Table~\ref{tab:judge}). The judge serves as a qualitative diagnostic only; the composite score in Table~\ref{tab:master} relies on the embedding cosine $C_\text{emb}$ rather than on the judge's ratings.
DeepSeek dominates all SA dimensions (4.04/5), while GPT-4.1 leads in DA mode (3.36/5). The largest SA--DA gap appears in the Result-Completeness dimension, consistent with the observation that the Dual Agent's replanning loop interrupts successful task completions on strong models.
\begin{table}[H]
\centering
\caption{LLM-as-Judge reasoning assessment (1--5 scale). Bold = best per architecture. The five dimensions capture task understanding, data handling, methodology, self-correction, and result completeness.}
\label{tab:judge}
\small
\resizebox{\columnwidth}{!}{%
\begin{tabular}{l ccccc c}
\toprule
\textbf{Model} & \textbf{Task} & \textbf{Data} & \textbf{Method} & \textbf{Self-Corr.} & \textbf{Result} & \textbf{Avg.} \\
\midrule
\multicolumn{7}{l}{\textit{Single Agent}} \\
DeepSeek     & 4.14 & \textbf{4.48} & \textbf{3.98} & \textbf{3.44} & \textbf{4.18} & \textbf{4.04} \\
Gemini Flash & \textbf{3.78} & 4.10 & 3.42 & 3.16 & 3.46 & 3.58 \\
GPT-5.4      & 3.28 & 3.98 & 3.14 & 3.12 & 3.12 & 3.33 \\
GPT-4.1      & 3.26 & 3.62 & 3.02 & 2.48 & 2.76 & 3.03 \\
Llama-70B    & 3.04 & 3.42 & 2.52 & 2.64 & 1.86 & 2.70 \\
Qwen-14B     & 2.81 & 2.94 & 2.23 & 2.30 & 1.66 & 2.39 \\
\midrule
\multicolumn{7}{l}{\textit{Dual Agent}} \\
GPT-4.1      & \textbf{3.66} & \textbf{3.48} & \textbf{3.18} & 3.46 & \textbf{3.04} & \textbf{3.36} \\
GPT-5.4      & 3.34 & 3.00 & 2.96 & \textbf{3.64} & 2.76 & 3.14 \\
Gemini Flash & 2.94 & 3.22 & 2.54 & 3.10 & 2.44 & 2.85 \\
Llama-70B    & 3.10 & 2.54 & 2.40 & 3.08 & 1.92 & 2.61 \\
DeepSeek     & 3.10 & 2.76 & 2.40 & 3.04 & 1.38 & 2.54 \\
Qwen-14B     & 2.92 & 2.31 & 2.24 & 2.86 & 1.53 & 2.37 \\
\bottomrule
\end{tabular}}
\end{table}
\FloatBarrier
\subsubsection{Output accuracy and composite scores (L3)}
\begin{figure}[!htbp]
\centering
\includegraphics[width=\columnwidth]{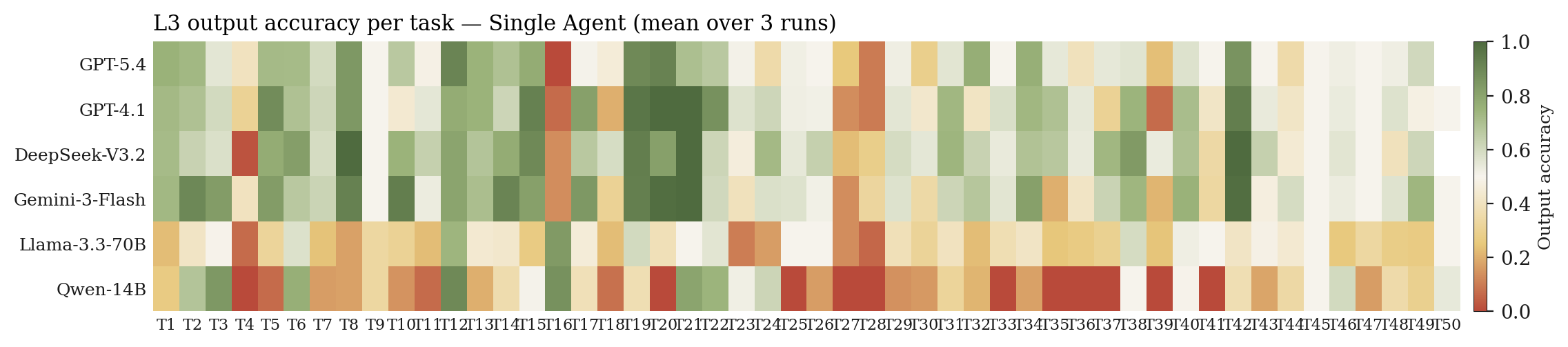}
\caption{L3: Per-task output accuracy scores (SA, averaged over three runs). Green = high accuracy; red = failure. Tasks T18 and T27 remain red across most models, indicating intrinsic difficulty.}
\label{fig:l3_heatmap}
\end{figure}

Fig.~\ref{fig:l3_heatmap} provides a per-task breakdown of L3 output accuracy scores across SA mode (mean over three runs).
The heatmap retains a bimodal pattern: most tasks cluster near $0.5$--$1.0$ or near $0$, with few intermediate values, indicating that task success is largely binary once threshold quality is achieved.
A quality-aware composite is therefore essential: many failed tasks complete more than $70\%$ of the required pipeline (correct spatial joins, Moran's~I computation, or GWR fitting), yet binary pass/fail scoring cannot capture this partial progress.
The vision-based L3 evaluator (Section~\ref{sec:eval}) further enforces a strict cartographic standard: $33\%$ of SA visualizations and $58\%$ of DA visualizations receive task-completion scores $\leq 2$, even when the underlying analytical pipeline ran to completion---typically due to missing legends, incorrect colour ramps, or extents that omit the gold-standard area of interest.
This strict cartographic gate is the principal driver of the per-task $Q_\text{out}$ values in Table~\ref{tab:master} sitting well below the corresponding success rates.
The composite scores in Table~\ref{tab:master} integrate all evaluation layers.
Gemini-3-Flash narrowly leads SA composite ($0.7646$), with DeepSeek-V3.2 a close second ($0.7612$); both maintain similar profiles in DA mode but at $5$--$15$ percentage-point lower composite values.
GPT-5.4 holds the lead in DA mode ($0.7052$), reflecting the planner-friendly nature of its longer-context reasoning.
\begin{figure}[!htbp]
\centering
\includegraphics[width=\columnwidth]{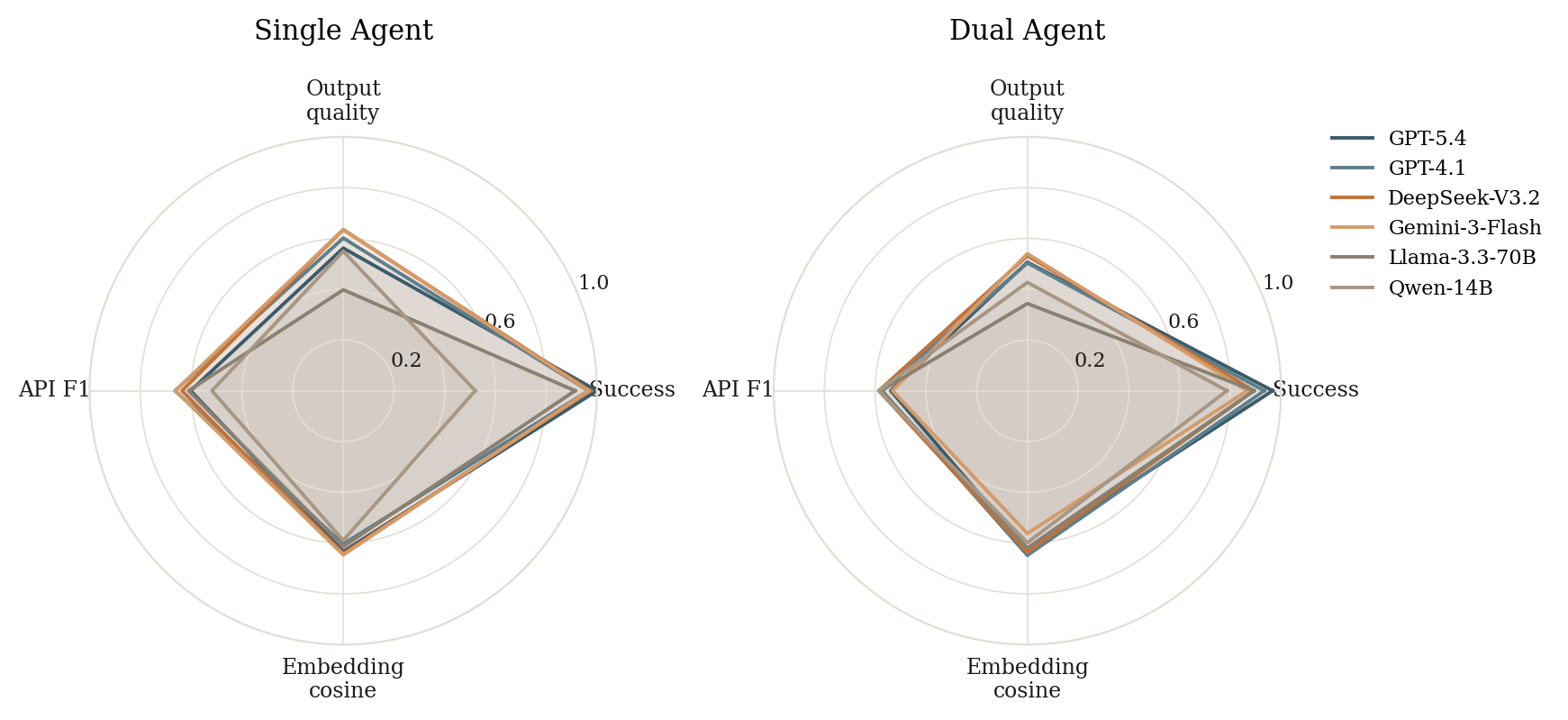}
\caption{Multi-dimensional performance profiles for SA (left) and DA (right) architectures. In SA mode, Gemini Flash and DeepSeek lead the four-dimensional composite; in DA mode, GPT-class models hold the highest composite but with compressed dimension ranges across all six backends.}
\label{fig:radar}
\end{figure}

\FloatBarrier
\subsubsection{Run-to-run variance: the case for repeated runs}\label{sec:variance}
A central methodological finding from the 1{,}800-experiment grid is that single-run task scores are surprisingly noisy, even for flagship models on tasks that are within their capability envelope.
The clearest example is T20 (mineral deposit prospectivity prediction), where DeepSeek-V3.2 in SA mode produces task-completion vision scores of $1$, $3$, and $5$ across the three runs---spanning the full failure-to-success range while every other strong model on T20 scores $5/5/5$ deterministically.
Any single execution of T20 would therefore class DeepSeek as either an unambiguous failure or an unambiguous success depending purely on the random branch sampled, and identical single-run instability appears at smaller magnitudes on $9$ further tasks.
Aggregated across $50$ tasks, this run-to-run noise tightens the bootstrap CIs in Table~\ref{tab:master} to $\pm 0.02$--$0.05$, but cell-level conclusions drawn from any single execution can shift by similar margins in either direction.
The three-run protocol is therefore not a cosmetic statistical addition; it is a precondition for stable architecture- and model-level rankings, and we recommend it as a baseline requirement for future GIS-agent benchmarking.

\FloatBarrier
\subsubsection{Effect of optional workflow hints}\label{sec:ablation}
Throughout the main evaluation we supply each task with the high-level workflow outline that the benchmark provides; this represents the typical use case in which a domain expert sketches the analytical decomposition for the agent.
To check how much this optional hint actually moves the needle---and how the system performs when no such hint is available---we run a paired ablation on DeepSeek-V3.2 SA across all 50 benchmark tasks, with vs.\ without the workflow string injected into the prompt (all other configuration held identical).
Table~\ref{tab:ablation} reports the result.
End-to-end task success is essentially unchanged ($47/50$ with workflow, $48/50$ without): the system reaches the analytical end of nearly every task either way.
The composite score, which additionally weighs output quality and process alignment, shows a moderate improvement when the workflow is supplied (mean $\Delta = +0.114$, median $\Delta = +0.153$, Cohen's $d = +0.51$, paired Wilcoxon $p < 10^{-3}$)---driven by tasks where the workflow helps the model select the expert-aligned analytical decomposition over plausible alternatives.
The takeaway is two-sided: an expert-supplied workflow is a useful but optional input that moderately sharpens the agent's outputs, and the system itself is robust enough to perform realistic GIS analysis even without one.
\begin{table}[H]
\centering
\caption{Workflow ablation: DeepSeek-V3.2 SA across all 50 benchmark tasks, paired by task ($n=50$). Composite scores follow Eq.~\ref{eq:composite}. The with-workflow condition delivers a statistically significant medium-sized improvement.}
\label{tab:ablation}
\small
\begin{tabular}{l rrr}
\toprule
\textbf{Condition} & $\boldsymbol{S_\textbf{comp}}$ \textbf{mean} & \textbf{std} & \textbf{successes} \\
\midrule
With workflow    & $0.733$ & $0.196$ & $47 / 50$ \\
Without workflow & $0.619$ & $0.180$ & $48 / 50$ \\
\midrule
\multicolumn{4}{l}{Mean per-task $\Delta$ (with $-$ without) $= +0.114$;\quad median $\Delta = +0.153$.} \\
\multicolumn{4}{l}{Paired Wilcoxon $p < 10^{-3}$;\quad Cohen's $d = +0.51$ (medium effect).} \\
\bottomrule
\end{tabular}
\end{table}
\begin{figure*}[!htbp]
\centering
\includegraphics[width=\textwidth]{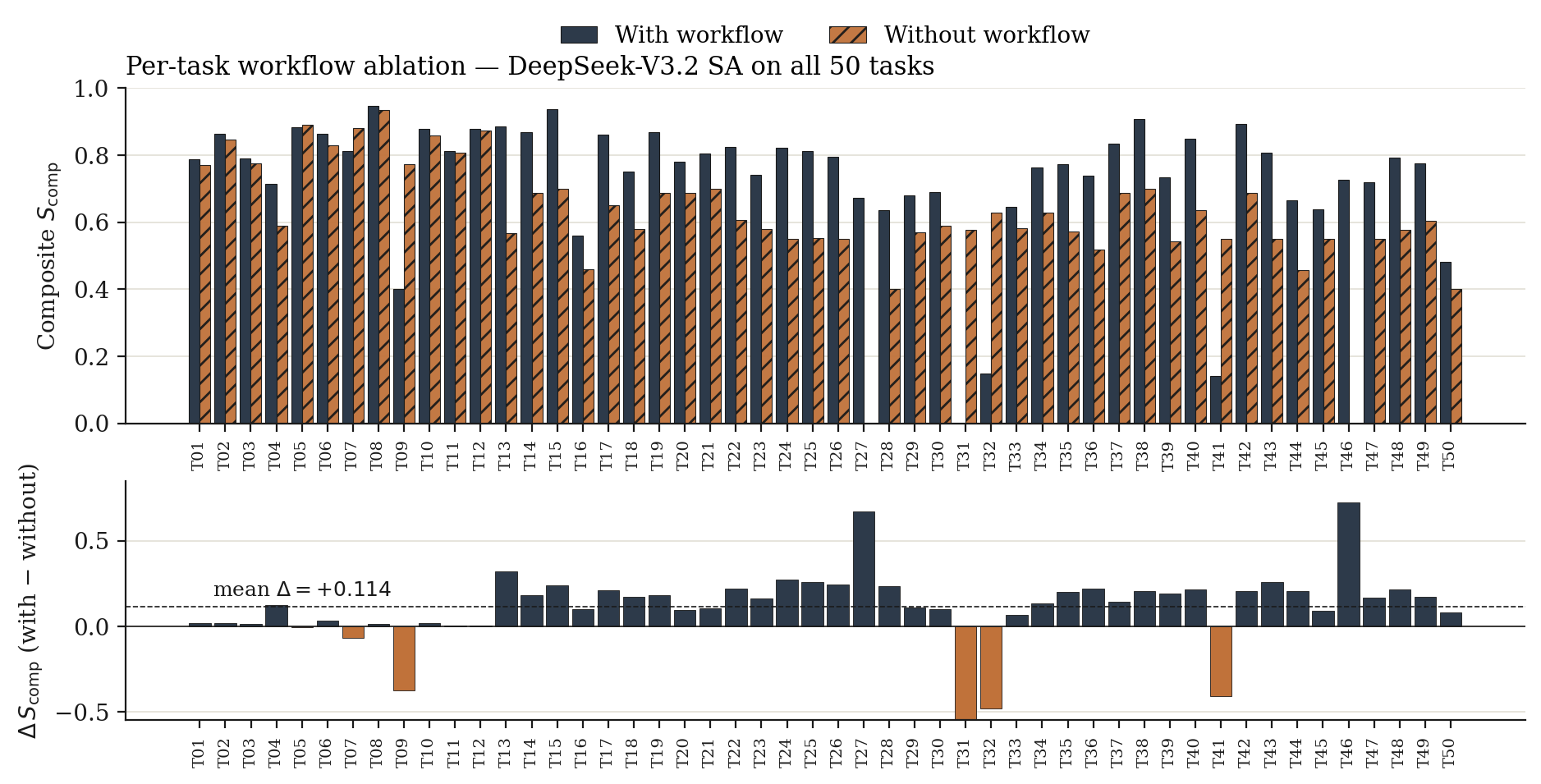}
\caption{Per-task workflow ablation across all 50 tasks (DeepSeek-V3.2 SA). Top: paired composite scores with vs.\ without workflow. Bottom: signed per-task difference $\Delta = S_\mathrm{comp}^\mathrm{with} - S_\mathrm{comp}^\mathrm{without}$ (slate = with-workflow wins, warm = without wins); the dashed line marks the mean $\Delta = +0.114$. The majority of tasks favour the with-workflow condition, with a small minority of reversals---most prominently T41, where the supplied workflow encodes a constraint that mismatches the model's preferred decomposition---outweighed by the consistent positive effect on multi-step pipelines.}
\label{fig:ablation_bar}
\end{figure*}
\FloatBarrier
\subsection{Case Studies}\label{sec:casestudies}
We present three categories of case studies that illuminate the interaction between model capability, domain knowledge, and system design.
\FloatBarrier
\subsubsection{Exceptional performance on complex tasks---and run-to-run instability}
\begin{figure*}[!htbp]
\centering
\includegraphics[width=\textwidth]{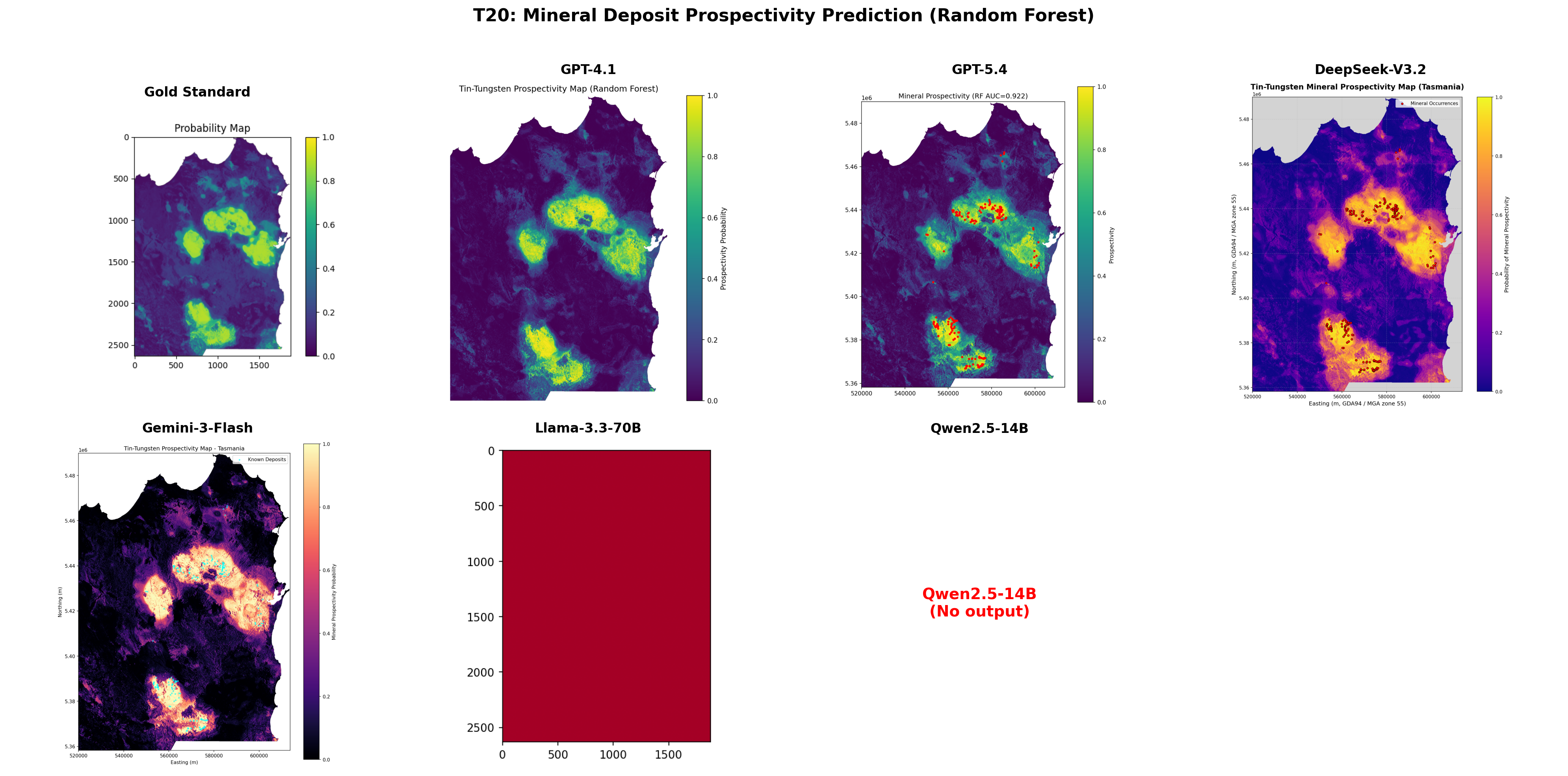}
\caption{T20: Mineral deposit prospectivity prediction outputs (SA mode). Four strong models accurately reproduce the spatial distribution of tin-tungsten prospectivity, while Llama-70B generates a uniform raster and Qwen-14B fails to produce any visualization. This task requires a complete ML pipeline spanning raster feature extraction, Random Forest training, and georeferenced probability mapping.}
\label{fig:case_t20}
\end{figure*}

\textbf{T20 --- Mineral deposit prospectivity prediction.}
This task requires a complete machine learning pipeline: loading 11 geological raster layers (geochemistry, geophysics, lithology, structure), extracting pixel-level features, training a Random Forest classifier on known tin-tungsten deposit locations, generating a full-resolution probability map, and visualizing the prospectivity surface.
The pipeline spans raster I/O, nodata handling, feature matrix construction, scikit-learn model fitting, pixel-wise prediction, and georeferenced output---a representative end-to-end ML-driven geospatial workflow.
As shown in Fig.~\ref{fig:case_t20}, four cloud models (GPT-4.1, GPT-5.4, DeepSeek-V3.2, Gemini-3-Flash) produce prospectivity maps that faithfully reproduce the spatial pattern of the gold standard, with high-probability zones correctly localized around known deposit clusters; Llama-3.3-70B generates a uniform single-color raster (complete pipeline failure), and Qwen-14B fails to produce a visualization.
GPT-5.4 additionally reports a model AUC of $0.922$ and overlays known deposit points, while DeepSeek annotates the coordinate reference system (GDA94 / MGA zone 55).

\begin{figure}[!htbp]
\centering
\includegraphics[width=\columnwidth]{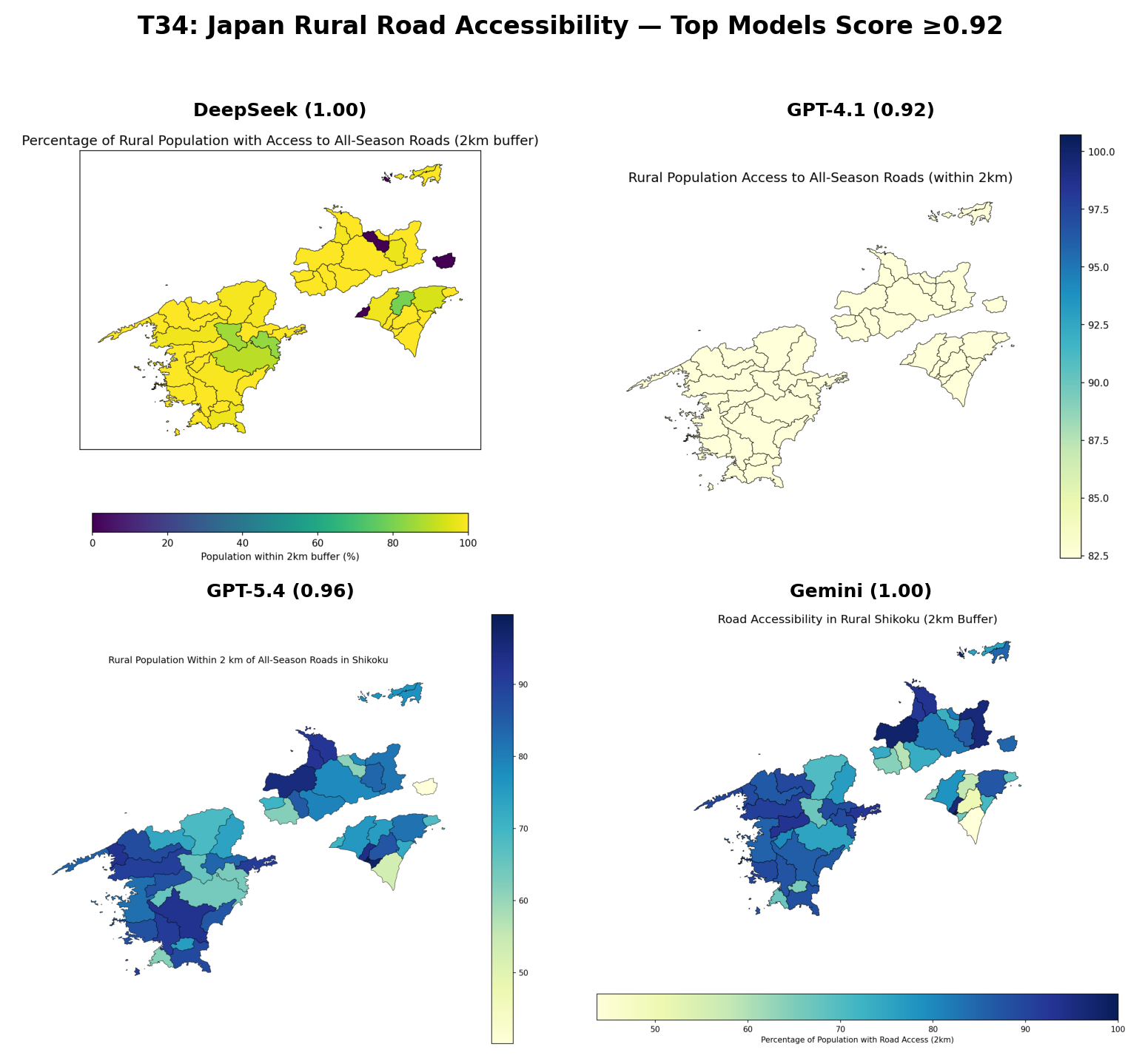}
\caption{T34: Road accessibility visualization outputs. Top models accurately identify and render prefecture-level accessibility patterns across Japan.}
\label{fig:case_t34}
\end{figure}

Critically, T20 also exposes the run-to-run instability discussed in Section~\ref{sec:variance}: the L3 vision task-completion score for DeepSeek SA on T20 is $1$, $3$, $5$ across the three repeated runs, whereas GPT-5.4 and GPT-4.1 score $5/5/5$ deterministically.
T20 is therefore simultaneously a positive case (strong models can orchestrate multi-step ML pipelines without explicit procedural guidance) and a cautionary case (high-temperature creative branches in flagship models occasionally redirect the pipeline through subtly wrong feature matrices), and is the single clearest empirical motivation for the repeated-run protocol.
\textbf{T34 --- Japan rural road accessibility analysis.}
This multi-step GIS pipeline requires rural area extraction, road buffer construction, population ratio computation, and choropleth visualization.
Four models score above 0.92, with DeepSeek and Gemini achieving perfect scores (1.00).
Fig.~\ref{fig:case_t34} shows that successful outputs faithfully reconstruct the spatial accessibility pattern across Japan's prefectures.
\FloatBarrier
\subsubsection{Parametric domain knowledge gaps}
\begin{figure*}[!htbp]
\centering
\includegraphics[width=\textwidth]{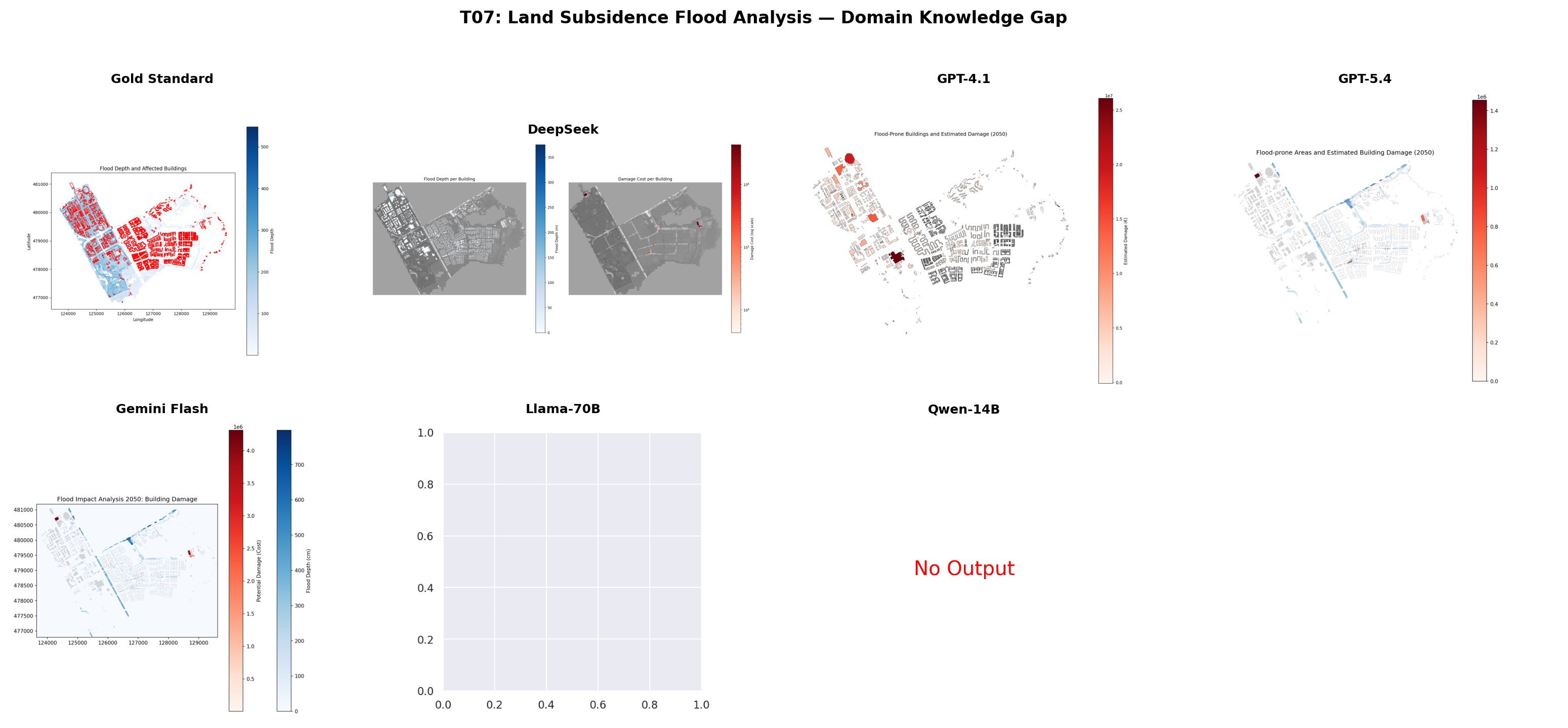}
\caption{T07: Flood analysis outputs illustrating domain knowledge gaps. The Gold Standard (left) uses a domain-calibrated $-200$\,cm threshold, while all models default to $0$\,cm. Some models further suffer CRS misalignment, producing blank maps. This demonstrates that \textit{parametric domain knowledge} cannot be acquired from code patterns alone.}
\label{fig:case_t07}
\end{figure*}

\begin{figure}[!htbp]
\centering
\includegraphics[width=\columnwidth]{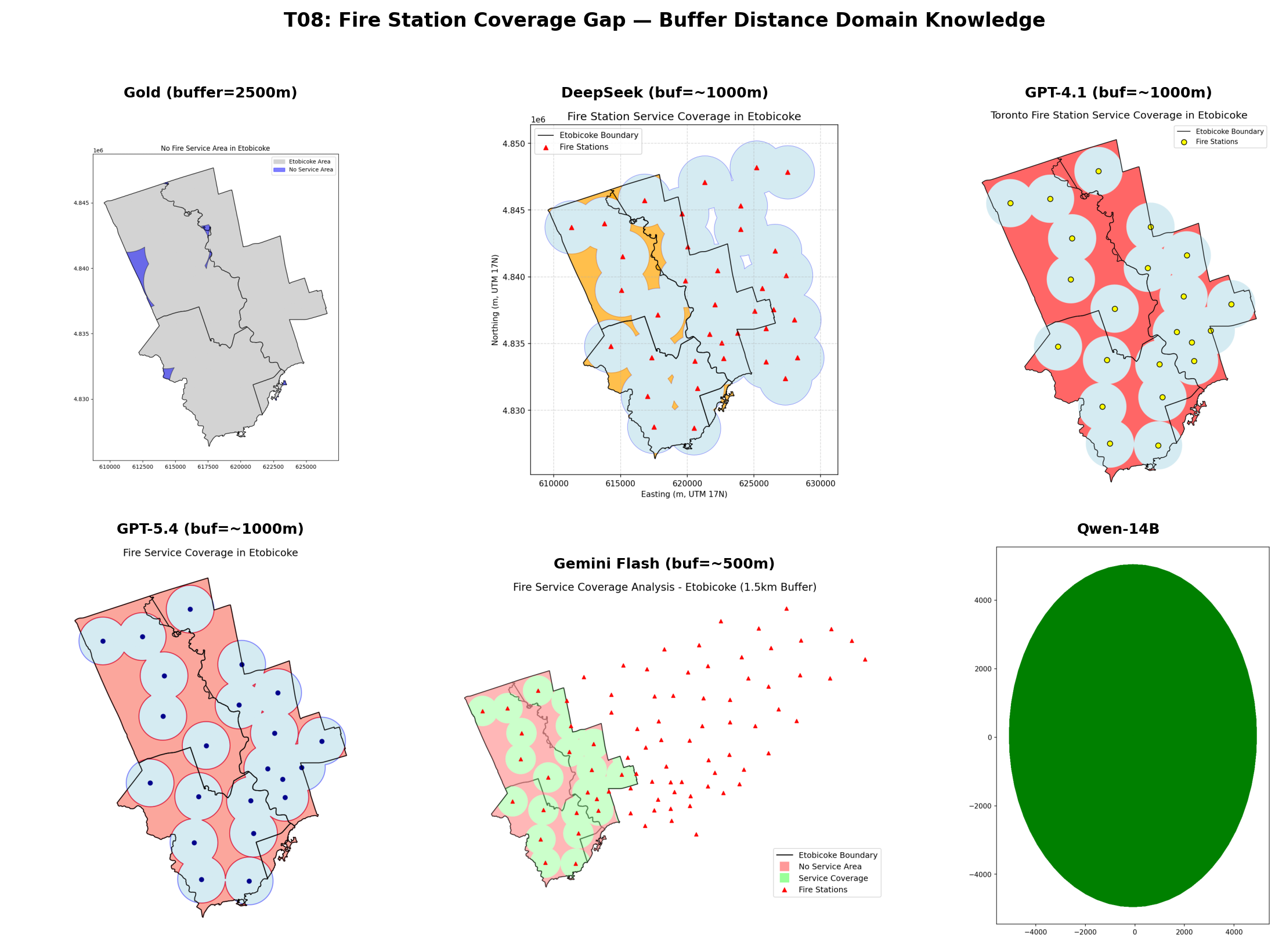}
\caption{T08: Fire station coverage gap analysis. The Gold Standard (top-left) uses a 2,500\,m buffer; all models select smaller radii (500--1,000\,m), dramatically overestimating the uncovered area. This illustrates the \textit{parametric domain knowledge gap}.}
\label{fig:case_t08}
\end{figure}

\textbf{T07 --- Land subsidence flood analysis.}
This task exposes a critical limitation: the gold standard uses a flood depth threshold of $\leq -200$\,cm, which encodes domain knowledge about projected future sea level rise.
All six models instead use the naive threshold \texttt{elevation < 0}, producing outputs that \textit{appear} correct but significantly overestimate flood extent.
Moreover, several models incorrectly hardcode the CRS as EPSG:4326 instead of the data's native EPSG:28992, causing the overlay operation to fail silently and producing near-blank visualizations (Fig.~\ref{fig:case_t07}).
\textbf{T08 --- Fire station coverage gap analysis.}
Similarly, the gold standard specifies a 2,500\,m service buffer based on fire response time standards, while models select arbitrary values (500--1,000\,m), producing ``no-service'' areas that are 3--5$\times$ larger than the reference (Fig.~\ref{fig:case_t08}).

These cases reveal a qualitative distinction that RAG-based systems cannot resolve: while RAG can supply library documentation, it cannot impart fundamental \textit{conceptual} domain knowledge (e.g., flood depth = sea level $-$ terrain elevation) or \textit{parametric} thresholds (e.g., fire-response buffer distances) that human analysts acquire through professional experience.
This ``conceptual knowledge gap'' represents a distinct failure mode from the ``API knowledge gap'' that retrieval-augmented systems were designed to address.

\FloatBarrier
\subsubsection{Infrastructure vs.\ model capability}
\textbf{T18 --- Quadtree density visualization.}
All models correctly identify \texttt{geoplot.quadtree()} as the appropriate function, but a compatibility issue between geoplot~0.5.1 and the installed pyproj version causes universal failure.
This case demonstrates that some ``model failures'' are actually \textit{environmental limitations}, underscoring the importance of infrastructure design described in the next section.
\FloatBarrier
\subsection{Design Lessons}\label{sec:infra_lessons}

Our $1{,}800$-experiment evaluation and iterative development process yield several engineering principles that extend beyond benchmark scores.

\FloatBarrier
\subsubsection{Infrastructure dominates model capability}
The primary bottleneck for GIS agent success is \textbf{infrastructure-level system design}, not model reasoning ability.
Through iterative failure diagnosis during system development, we resolved four categories of environment-level bugs (Table~\ref{tab:infra_fixes}), improving intermediate-task success from 20\% to 80\%---a 400\% increase---while simple and complex tasks remained unchanged.
This demonstrates that \textbf{the design bottleneck lies at the interface between model output and execution environment}.

\begin{table}[H]
\centering
\caption{Infrastructure-level fixes and their impact on task success.}
\label{tab:infra_fixes}
\small
\begin{tabular}{p{1.6cm} p{2.8cm} p{2.8cm}}
\toprule
\textbf{Category} & \textbf{Failure Symptom} & \textbf{Fix Applied} \\
\midrule
Data Pathing & Nested SHP files cause \texttt{FileNotFoundError} & Flatten dataset structure \\
\midrule
API State & \texttt{rasterio.open()} returns reader object; agent calls \texttt{.astype()} & Return raw \texttt{ndarray} + metadata dict \\
\midrule
Memory & City-wide grids exceed $10^4$ cells; \texttt{overlay} OOM & Enforce grid-size templates \\
\midrule
Type Coercion & String-encoded numerics fail during interpolation & Auto-cast with \texttt{pd.to\_numeric} \\
\bottomrule
\end{tabular}
\end{table}

Beyond bug fixes, tool layers can introduce \textit{silent semantic mismatches}.
In T36 (SAVI computation), a \texttt{load\_raster} helper returned a 2D array instead of a 3D $(B,H,W)$ array; the agent interpreted \texttt{data[3]} as a spectral band but retrieved a pixel row---producing wrong values with no exception.
GPT-5.4 spent 35 rounds diagnosing ``wrong values'' without locating the root cause; after fixing the interface, the same model succeeded in 3 rounds.

Two additional behavioral failures emerged: (1)~in T06, Qwen-14B called \texttt{plt.show()} instead of \texttt{plt.savefig()}, producing empty outputs---a single prompt rule resolved it (score: 3.5$\to$7.8); (2)~across 50 tasks, 8 failures (16\%) stemmed from file-name case ``auto-correction'' on case-sensitive Linux, eliminated by the Schema Analysis rule mandating \texttt{list\_files()} as the first action.

\FloatBarrier
\subsubsection{Hard constraints outperform prompt-level guidance}
Small models exhibit \textit{frustration-induced constraint collapse}: after repeated failures, they abandon prompt constraints and revert to pre-training defaults.
In T26 (groundwater mapping), Qwen-14B attempted \texttt{import arcpy} at round~17 despite explicit prohibition.
Comparing three strategies: prompt-only (blocked at round~17), prompt+deduplication (blocked earlier at round~10), and prompt+sandbox-level import interception (zero violations, full completion).
The lesson: \textbf{sandbox-level enforcement is a necessary complement to prompt-level constraints} for resource-constrained models.

A related finding: code deduplication prevents ``progress amnesia'' loops (\eg, T25: 8$\times$ identical submissions) but can prematurely terminate legitimate retries of slow operations (\eg, T28: \texttt{GridSearchCV} timeouts), suggesting that \textbf{context-aware deduplication} is a worthwhile future direction.

\FloatBarrier
\subsubsection{Two failure modes of self-correction}
Our analysis reveals two distinct self-correction failures.
\textit{Type~A (Conceptual):} the model lacks domain knowledge---in T33, it repeatedly applies the wrong flood-depth formula because the error is conceptual, not syntactic.
\textit{Type~B (Attentional):} the model correctly diagnoses the issue but fails to apply the fix---in T34, \texttt{print()} revealed \texttt{'RURAL'} but the next submission still used \texttt{'Rural'}.
Type~A requires domain knowledge injection with physical parameter specifications; Type~B may benefit from explicit memory prompting that foregrounds extracted values.

A related failure is \textbf{task instruction ambiguity}: in T25, ``create risk maps'' was interpreted as GeoJSON export rather than PNG visualization---a GIS-specific semantic convention that models cannot infer without explicit output-format contracts.

\FloatBarrier
\subsubsection{Local deployment viability and model selection}
Averaged across three runs, Qwen2.5-Coder-14B on a single RTX~3090 achieves $52\%$ end-to-end success in SA mode with output quality $Q_\text{out} = 0.549$, comparable to GPT-4.1's $0.601$ and well above Llama-3.3-70B's $0.398$.
Counter-intuitively, Llama-70B---with $5\times$ more parameters---scores lower on the SA composite ($0.656$ vs.\ Qwen-14B's $0.454$ once success rate is included, but with much weaker output quality), revealing a comprehension--execution gap: high embedding similarity ($0.614$) but persistently low output quality.
Equally striking, switching Qwen2.5-Coder-14B from SA to DA mode lifts mean success from $52\%$ to $79\%$ ($+27$\,pp) and composite from $0.454$ to $0.588$, the only such gain across the six backends.
\textbf{For local deployment, code-specialized fine-tuning is more impactful than parameter scaling, and the additional planner scaffolding is most valuable precisely where single-shot reasoning is weakest.}

Qualitative analysis shows that Qwen-14B failures concentrate in the \textit{final output assembly stage}, not in core GIS reasoning.
In T34, T01, and T20, the model correctly executed spatial joins, Kriging preparation, and Random Forest training, but failed at visualization or GeoTIFF reassembly.
This suggests a \textbf{human-in-the-loop} paradigm where the agent handles analytical stages (estimated $>$80\% completion) while the expert validates at key transitions---a pragmatic path to capable geospatial AI on consumer hardware.
\FloatBarrier
\section{Discussion and Limitations}\label{sec:limitations}
\textbf{Scope of agent autonomy.}
Each task in the main benchmark is accompanied by an optional high-level workflow outline (a short ordered list of analytical steps such as \textit{interpolate temperatures, then filter senior population, then aggregate to census blocks}).
The workflow specifies neither code, function names, parameter values, coordinate systems, nor output formats; the agent is responsible for translating the natural-language description into executable Python, handling all data realities (column names, CRS, missing values), recovering from runtime errors, and producing the final analytical outputs.
We use this hint by default because it is what a real GIS analyst would typically supply when delegating an analysis, and not because the system requires it: the controlled ablation in Section~\ref{sec:ablation} shows that removing the workflow on a strong cloud backend leaves task success essentially unchanged ($48/50$ vs.\ $47/50$), with only a moderate composite-score reduction ($\Delta = -0.114$, $d = -0.51$) attributable to looser alignment with the expert decomposition.
\sys{} should therefore be read as a system that performs realistic GIS analysis in a setting that mirrors how human analysts actually use such systems---with a brief expert sketch when one is available, and with self-directed decomposition when one is not---rather than as a claim of fully autonomous geospatial reasoning.

\textbf{Statistical scope.}
Each $(\text{model}, \text{architecture}, \text{task})$ combination in the main comparison was executed three times under identical sampling configurations, yielding $n = 150$ paired task-runs per (model, architecture) cell; we report means with $95\%$ confidence intervals computed by the percentile bootstrap (typical CI half-width $\pm 0.02$--$0.05$ on the composite score).
Architecture-level comparisons are tested with task-level paired Wilcoxon signed-rank tests (Table~\ref{tab:wilcoxon}), and effect sizes are reported using Cliff's~$\delta$.
We additionally conduct a sensitivity analysis sweeping all $66$ admissible weight tuples on the composite-score simplex, confirming that the multi-architecture and multi-model conclusions are not artefacts of any specific weight choice (median Kendall's $\tau = 0.94$ vs.\ the default weights).
Despite these safeguards, three runs per setting remain a moderate sample, and broader claims about absolute success rates---particularly in the long tail of difficult tasks---should be read with the reported uncertainties in mind.
The optional-workflow ablation in Section~\ref{sec:ablation} covers DeepSeek-V3.2 SA on all 50 benchmark tasks; extending the same ablation across the remaining five backends is a natural follow-up.

\textbf{Generalisation beyond \bench{}.}
\bench{} contains 50 expert-curated tasks spanning seven analytical categories.
While these tasks cover the major modalities (vector, raster, tabular) and a range of difficulty, they are predominantly drawn from urban analytics, environmental hazard, and resource-planning domains.
Performance on tasks demanding very large raster processing ($>$10~GB), real-time spatial data streams, multi-user collaborative analysis, or deeply specialised sub-domains (e.g.\ marine acoustics, atmospheric chemistry) is not assessed here, and the qualitative findings around dual-agent degradation and infrastructure-bound failure modes should be re-validated in those settings before being relied upon.

\textbf{Model availability and reproducibility.}
Several of the cloud models evaluated in this work---most notably the Gemini~3 Flash preview---are vendor-managed services whose underlying weights, sampling defaults, and quotas may change without notice.
We document exact model identifiers, prompt templates, and sandbox configurations to maximise reproducibility, but exact replication of cloud results may become infeasible as services evolve.
The open-weight component of our evaluation (Qwen2.5-Coder-14B, Llama-3.3-70B) is fully reproducible from public artefacts.

\textbf{Cost and ergonomics.}
Strong cloud models incur non-trivial inference cost (GPT-5.4 at approximately \$8.50 per full benchmark sweep), making large-scale repeated evaluation expensive.
The 14B open-weight tier mitigates this for routine use, but at the cost of slower per-task wall time and lower performance on the most complex tasks.
A practical pipeline therefore likely combines cheap open-weight models for routine workflows with selective routing to a stronger cloud model for tasks that the agent or its evaluator flags as ambiguous.

\FloatBarrier
\section{Conclusion}\label{sec:conclusion}
We presented \sys{}, a comprehensive open-source LLM agent system that performs realistic, multi-step GIS analysis---kriging-based hot-spot mapping, Random-Forest mineral prospectivity prediction, multi-criteria wildlife corridor optimization, network-based isochrones, full publication-quality cartography---directly through the open-source Python ecosystem, with no dependency on any commercial GIS software.
By grounding agent behaviour in a persistent Python sandbox, three engineered prompt rules (Schema Analysis, Package Constraint, optional Domain Knowledge Injection), and an Error-Memory module for self-correction, \sys{} reaches up to $100\%$ task success on \bench{} (mean $97\%$ over three independent runs for the leading SA backends) across the full range of vector, raster, and tabular workflows that prior LLM-driven GIS systems do not jointly support.

\textbf{Realistic GIS analysis without proprietary GIS.}
\sys{} demonstrates that the spectrum of professional analytical pipelines normally addressed with ArcGIS or QGIS---spatial overlays, raster algebra, geostatistical interpolation, machine-learning classification, network analysis, multi-layer cartography---can be performed end to end on the open-source Python stack alone, when paired with an appropriately engineered LLM agent.
The benchmark used here, \bench{}, is itself a deliberate stress test: tasks average $5.8$ analytical steps, span all three data modalities, and were authored by domain experts rather than synthesised from tool-call templates, which makes the headline success rates a meaningful indicator of system capability rather than of benchmark gaming.
The system, the prompt templates, and our open-source-rewritten gold-standard solutions for all 50 tasks are released under MIT licence to support direct use, replication, and extension.

\textbf{Methodological findings.}
The $1{,}800$-experiment evaluation grid supports two methodological points that we believe will be useful to the broader GIS-agent community.
First, per-task scoring exhibits non-trivial run-to-run variance even for flagship models (a single execution of T20 can score $1$, $3$, or $5$ on visual task completion across runs of an otherwise stable strong model), so a repeated-run protocol with confidence intervals and paired tests should be a default for GIS-agent benchmarking rather than an optional extra.
Second, within \sys{}'s identical infrastructure the Single-Agent ReAct loop reliably outperforms the Dual-Agent Plan-Execute-Replan pipeline on every cloud backend (paired Wilcoxon $p < 10^{-3}$, Cliff's $\delta = 0.21$--$0.41$), with only the locally deployed 14B model gaining from multi-agent orchestration; the implication is that architectural complexity should be matched to model capability rather than added by default.

\textbf{Infrastructure as the primary bottleneck.}
Our iterative failure diagnosis reveals that the performance ceiling for GIS agents is set not by model reasoning quality but by infrastructure design.
Resolving four categories of environment-level bugs---data pathing, API state management, memory limits, and type coercion---produced a 400\% improvement in intermediate-task success.
Additional failure modes at the behavioral layer---visualization output format, file-name guessing, and code deduplication side effects---further demonstrate that a system's engineering quality is a first-class determinant of agent performance, independent of the underlying model.

\textbf{The knowledge gap hierarchy.}
Our evaluation identifies a hierarchy of knowledge gaps that GIS agents face.
The \textit{API knowledge gap}---unfamiliarity with specific library calls---is addressable through prompt engineering and retrieval augmentation.
The \textit{parametric knowledge gap}---domain-calibrated thresholds such as flood depth cutoffs and fire-station buffer radii---is not recoverable from code corpora and produces ``false positive'' outputs that appear plausible but are analytically incorrect.
The \textit{conceptual knowledge gap}---physical relationships such as $\text{flood depth} = -(\text{elevation} + 200)$, and GIS-specific semantic conventions such as ``create maps'' implying PNG visualization and not GeoJSON export---lies beyond what any retrieval system can supply and calls for expert-in-the-loop validation or output format contracts baked into domain knowledge injection.

\textbf{Local deployment and human-in-the-loop pathways.}
Averaged across three repeated runs, Qwen2.5-Coder-14B on a single consumer RTX~3090 achieves $52\%$ end-to-end task success in SA mode (rising to $79\%$ in DA mode) at zero marginal cost, with output quality among successful tasks ($Q_\text{out} = 0.549$) comparable to GPT-4.1 ($0.601$).
Qualitative analysis shows that failures concentrate in final output assembly rather than in core GIS reasoning or spatial computation---indicating that the gap between 14B models and full autonomy is narrower than aggregate metrics suggest.
For deployments where cloud APIs are unavailable, a human-in-the-loop collaborative mode---agent handles pipeline stages autonomously while the domain expert validates at key transitions---offers a pragmatic path to capable geospatial AI without proprietary infrastructure.

\textbf{Future directions.}
Near-term priorities include extending \sys{} toward real-time geospatial data streams, incorporating context-aware deduplication, and adapting agent frameworks for reasoning-specialized models that require expanded token budgets.
Longer-term, the conceptual and parametric knowledge gaps identified in this work motivate hybrid architectures coupling LLM code generation with domain knowledge bases that encode field-specific physical models and calibrated parameter ranges---advancing GIS agents from syntactic proficiency toward genuine geoscientific reasoning.
\section*{Acknowledgements}
This study was supported by a grant from the National Research Foundation, Korea (NRF), funded by the Korean government (MSIT) (RS-2026-25474388).
This research was supported by the 2023-MOIS36-004 (RS-2023-00248092) of the Technology Development Program on Disaster Restoration Capacity Building and Strengthening funded by the Ministry of Interior and Safety (MOIS, Korea).
\section*{Data Availability}
\sys{}, including all source code, evaluation scripts, and benchmark results, is publicly available at: \url{https://github.com/geumjin99/GISclaw}.
\bibliographystyle{plainnat}
\bibliography{references}
\end{document}